\documentclass[%
 reprint,
superscriptaddress,
amsmath,amssymb,
aps,
pre,
showpacs,
longbibliography
]{revtex4-1}

\usepackage[utf8]{inputenc}
\usepackage{graphicx}
\usepackage{dcolumn}
\usepackage{bm}
\usepackage{tikz}
\usepackage{tabularx}
\usepackage{float}
\usepackage{amssymb}
\usepackage{pgfplots}
\pgfplotsset{compat=1.3}
\usetikzlibrary{plotmarks}
\usepackage{rotating}
\usepackage{hyperref}
\newcommand{\tg}[1]{\textcolor{black}{ #1}}
\newcommand{\seb}[1]{\textcolor{black}{ #1}}

\begin{document}
\title{Mechanics and structure of carbon black gels under high-power ultrasound}
\author{Noémie Dagès}
\thanks{Equal contribution}
\affiliation{Univ Lyon, Ens de Lyon, Univ Claude Bernard, CNRS, Laboratoire de Physique, F-69342 Lyon, France}%
\author{Pierre Lidon}
\thanks{Equal contribution}
\affiliation{Univ Lyon, Ens de Lyon, Univ Claude Bernard, CNRS, Laboratoire de Physique, F-69342 Lyon, France}%
\author{Guillaume Jung}
\affiliation{Univ Lyon, Ens de Lyon, Univ Claude Bernard, CNRS, Laboratoire de Physique, F-69342 Lyon, France}%
\author{Frédéric Pignon}
\affiliation{Univ Grenoble Alpes, CNRS, LRP, F-38000 Grenoble, France}%
\author{Sébastien Manneville}
\affiliation{Univ Lyon, Ens de Lyon, Univ Claude Bernard, CNRS, Laboratoire de Physique, F-69342 Lyon, France}%
\author{Thomas Gibaud}
\email{Corresponding author, thomas.gibaud@ens-lyon.fr}
\affiliation{Univ Lyon, Ens de Lyon, Univ Claude Bernard, CNRS, Laboratoire de Physique, F-69342 Lyon, France}%

\date{\today}

\begin{abstract}
Colloidal gels made of carbon black particles \tg{dispersed in light mineral oil} are ``rheo-acoustic'' materials, i.e., their mechanical and structural properties can be tuned using high-power ultrasound, sound waves with submicron amplitude and frequency larger than 20~kHz . The effects of high-power ultrasound on the carbon black gel are demonstrated using two experiments: rheology coupled to ultrasound to test for the gel mechanical response and a timeresolved ultra small-angle X-ray scattering experiment (TRUSAXS) coupled to ultrasound to test for structural changes within the gel. We show that high-power ultrasound above a critical amplitude \tg{leads to a complex viscoelastic transient response of the gels within a few seconds: a softening of its storage modulus accompanied by a strong overshoot in its loss modulus. Under high-power ultrasound, the gel displays a viscoelastic spectrum with glass-like features and a significant decrease in its yield strain. Those effects are attributed to the formation of intermittent micro-cracks in the bulk of the gel as evidenced by TRUSAXS.}
\tg{Provided that the shear rate is not large enough to fully fluidize the sample, high-power ultrasound also facilitates} the flow of the gel, \tg{reducing its yield stress as well as increasing the shear-thinning index}, thanks again to the formation of micro-cracks. 
\end{abstract}

\maketitle

\section{Introduction}
A colloidal gel forms as attractive colloidal particles dispersed in a fluid aggregate into a space-spanning network~\cite{Larson:1999,Mewis:2012,Gibaud2020} leading to peculiar properties. Indeed, gels behave like soft elastic solids at rest and easily flow upon application of mild external stresses including shear, compression, gravity or vibrations. Those properties are of prime importance for applications to material design and industrial processes \cite{Balmforth:2014,Bonn:2017} in construction materials like cement \cite{lootens2004,Ioannidou:2016}, in food science~\cite{Mezzenga:2005,gibaud2012}, as well as in ink-jet printing \cite{lewis2002,smay2002,Tan:2018} or flow-cell batteries \cite{youssry2013,Fan:2014,Helal:2016,Narayanan:2017}. 

In practice, however, applying a mechanical strain or stress in order to disrupt the gel at will is far from ideal as it involves stirring devices such as pumps, motors or other rotating tools that may not be compatible with applications. In particular, in industrial lines dedicated to particle fabrication, gels may form in the process and clog the production pipe. In such a case, the production must be stopped and the pipe cleaned. Using high-power ultrasound could be very useful to fluidize the gels ex-situ, without dismounting the line. That is why we turn our attention to high-power ultrasound as a means to externally trigger mechanical changes in the gels. 


 \begin{figure*}
 	\centering
 	\includegraphics[width=0.95\textwidth]{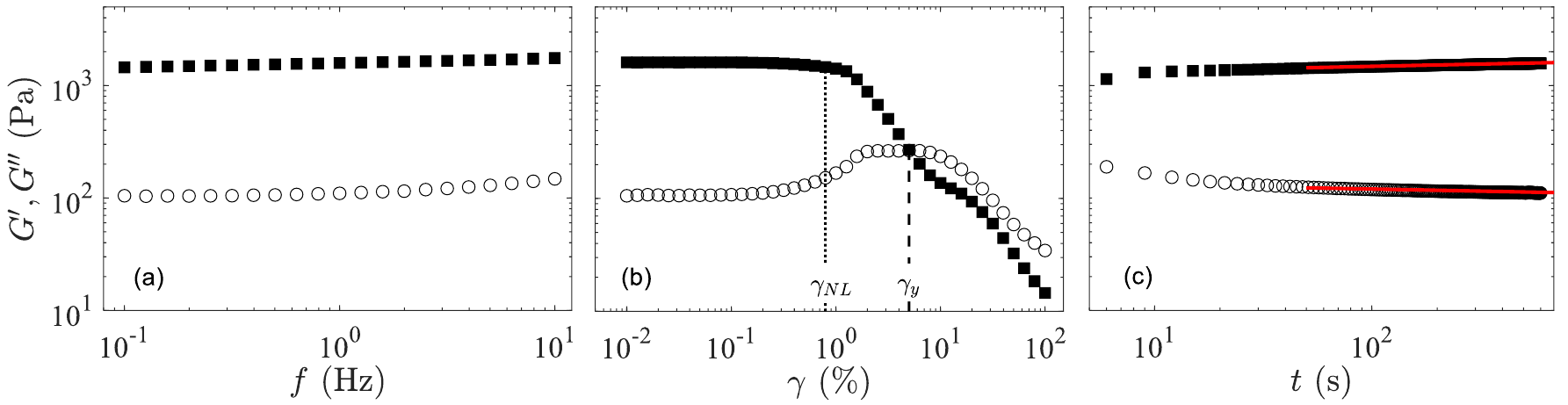}
         \caption{{\bf Viscoelastic properties of a 6\%~w/w carbon black gel.} Storage modulus $G'$ (full squares) and loss modulus $G''$ (empty circles) as a function of (a) frequency $f$ in a small-amplitude oscillatory shear at a fixed strain amplitude $\gamma=0.1\%$ and (b) strain amplitude $\gamma$ at a fixed frequency of $f=1~$Hz with a waiting time of $17~$s per point. Both $G'$ and $G''$ remain \tg{approximately} independent of the oscillatory strain amplitude $\gamma$ in the linear domain up to $\gamma_{NL} = 0.8$\% when $G'(\gamma_{NL})=0.85 G'(\gamma<0.1\%)$. The yield point is reached at $\gamma_y=5$\% when $G'=G''$. (c) Aging: $G'$ and $G''$ in the linear regime ($\gamma=0.1$\% and $f=1~$Hz) as a function of time $t$ after the preshear protocol. The red lines are logarithmic fits of the aging process. These experiments are performed in the parallel-plate geometry.
        }
     \label{fig:caract_cb}
 \end{figure*}

This possibility has recently been demonstrated on ``rheo-acoustic'' colloidal gels, gels sensitive to high-power ultrasound, a sound wave with submicron amplitude and frequency 20--500~kHz~\cite{gibaud2019}. \tg{In our previous study~\cite{gibaud2019}, ``rheo-acoustic'' effects were demonstrated on three different gel systems including carbon black, silica and calcite gels. We showed that \seb{ultrasound of high power, but below the intensity that induces acoustic cavitation,} significantly alters the gel structure and mechanics when the amplitude of the high-frequency oscillatory strain generated  by acoustic vibrations $\gamma_{US}$ becomes comparable to the strain $\gamma_{NL}$ above which the gel experiences nonlinear viscoelastic response under low-frequency oscillatory shear. 
Moreover, structural measurements showed that high-power ultrasound induces the formation of micro-cracks within the gels that softens gels at rest and facilitates their flow when a shear stress or rate is applied}. The goal of the present paper is to further explore the effects of high-power ultrasound on the structure and rheology of colloidal gels by focusing on carbon black gels. \tg{In particular, we aim at answering the following questions: What is the typical response time of the gel mechanical properties to high-power ultrasound? Beside a softening of the gel storage modulus $G'$, how does ultrasound affect the loss modulus $G''$ and, more generally, the gel viscoelastic spectrum and its nonlinear viscoelasticity? Can we quantify the dynamics of the micro-cracks identified in the previous study and their dependence on the ultrasound amplitude? How does flow compete with high-power ultrasound when both are combined?}

\tg{In the following, we} first describe the properties of \tg{ carbon black gels} at rest as well as the two setups used to impose ultrasound, namely a rheology experiment coupled to ultrasound to test the gel mechanical response and a Time-Resolved Ultra Small Angle X-ray Scattering experiment (TRUSAXS) coupled to ultrasound to test for structural changes in the gel. Second, we \tg{confirm} that carbon black gels are only sensitive to ultrasound above a threshold \tg{ultrasonic pressure}. 
\tg{Within a few seconds, high-power ultrasound leads to a decrease of the storage modulus while it enhances the loss modulus of the gel. We further show that, once a steady state is reached in the presence of ultrasound, the viscoelastic spectrum shows an upturn of the loss modulus at low frequencies that suggests that the microstructure becomes ``glass-like.'' This change goes along with a sharp decrease in the yield strain of the material.}
\tg{Thanks to complementary TRUSAXS experiments, we also deepen our understanding of the formation dynamics of \tg{intermittent} micro-cracks within the bulk of the gel due to ultrasound: we show that the occurrence frequency of intermittent micro-cracks strongly increases with the ultrasound amplitude and that, for large amplitudes, structural effects of ultrasound may persist long after vibrations are turned off.}
Finally, we focus on the effect of ultrasound on the flow properties of the gel by applying a constant shear rate on top of ultrasound. We show that the flow is \tg{facilitated, in the sense that the yield stress is lowered and that the shear-thinning index increases,} \seb{concomitantly with an increase in the occurrence of microcracks.}
\tg{This mechanism is dominant under flow at low shear rates but it gets less significant at the highest shear rates where the effects of ultrasonic vibrations are overcome by the flow-induced fluidization of the gel.}

 \begin{figure}[hbt!]
 	\centering
 	\includegraphics[width=0.95\columnwidth]{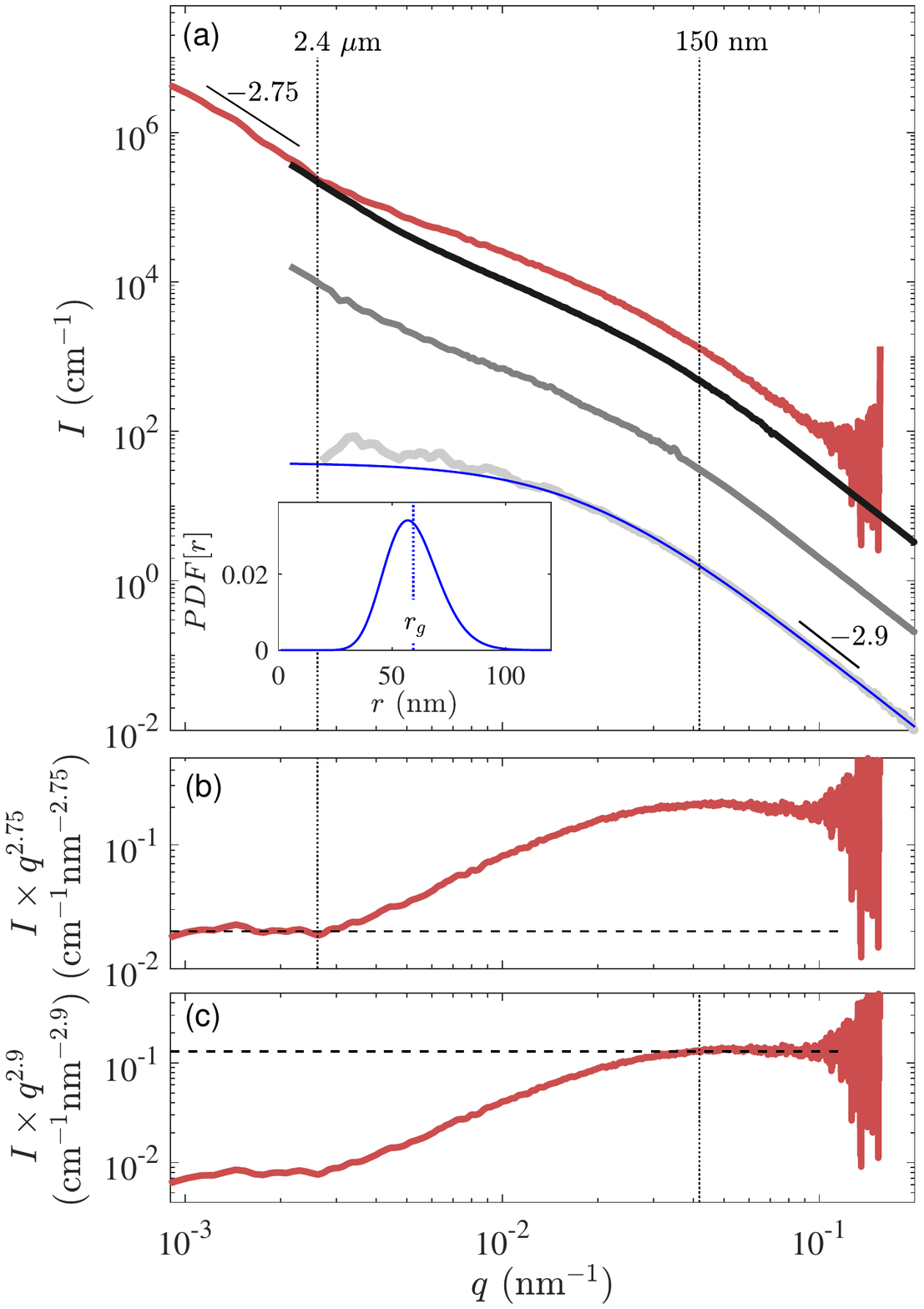}
         \caption{{\bf Scattering intensity from carbon black at different concentrations $c_w$ dispersed in mineral oil}. (a) SAXS intensity $I$ as a function of the magnitude $q$ of the scattering wave vector at $c_w=6$ (red), 1 (black), 0.1 (grey) and 0.01\%~w/w (light grey) from top to bottom. The blue line is a mass fractal fit of the form factor using a Schulz particle radius distribution centered on a radius of gyration $r_g=60$~nm with a polydispersity of 20\% as shown in the inset. (b)~Kratky-like representation for $c_w=6$\%~w/w highlighting the cut-off of the fractal structure at low $q$. (c)~Kratky-like representation for $c_w=6$\%~w/w highlighting the cut-off in the particle form factor  at high $q$. 
         }
     \label{fig:saxscb}
 \end{figure}

\section{Materials and methods}

\subsection{Carbon black gels, preparation, rheology and structure}

Carbon black gels are composed of colloidal carbon black particles (Cabot Vulcan XC72R, density of 1800~kg\,m$^{-3}$) dispersed in a mineral oil (density 838~kg m$^{-3}$, viscosity 20~mPa\,s, Sigma Aldrich) \cite{Gibaud:2010, Trappe:2007}. \seb{To disperse the carbon black in oil homogeneously, we mix a total mass of 50~g of oil and carbon black in a glass bottle of 100~mL. We first shake the bottle vigorously by hand and then place the bottle in a sonication bath (DK Sonic model DK-3000S with frequency 33.4~kHz and power 120~W) for one hour at maximum power}. In a mineral oil, the carbon black particles aggregate via van der Waals attraction \cite{Hartley:1985} and form a space-spanning network, a gel. Here, we focus on a dispersion of carbon black particles at a weight concentration $c_w=6\%$~w/w, which amounts to a volume fraction of 3\%. Due to the fractal nature of the particles, this corresponds to an effective volume fraction of about 20\% \cite{Trappe:2007}. Carbon black gels are \tg{known to be very sensitive} to the preshear history, which can be used to tune the initial microstructure of carbon black gels \cite{Osuji:2008b,Ovarlez:2013,Helal:2016,Richards:2017}. To ensure a reproducible initial state, we always use the same preshear protocol \tg{in our rheological experiments}. The preshear protocol is made of two successive steps of 200~s each, a first step at $-1000$~s$^{-1}$ in the ``reverse'' direction followed by one step at $+1000$~s$^{-1}$ in the ``positive'' direction, i.e., the direction of positive strains and stresses, in which shear will be applied during flow curve measurements in section~\ref{s:flowcurve}. After preshear, the dispersion recovers an elastic modulus within a few seconds. At low frequency $f\lesssim 10$~Hz, such a gel has a typical storage modulus $G'\simeq 10^3$~Pa and a loss modulus $G''\simeq 10^2$~Pa and shows little aging, see Fig.~\ref{fig:caract_cb}. \tg{Its viscoelastic response departs from linearity above a strain $\gamma_{NL}\simeq 0.8\%$ and the gel yields above a strain $\gamma_y\simeq 5\%$}. In addition to viscoelasticity, carbon black gels display peculiar mechanical and flow properties due to strong time-dependence under shear, including rheopexy \cite{Ovarlez:2013,Helal:2016,Hipp:2019}, delayed yielding \cite{Gibaud:2010,Grenard:2014} and fatigue \cite{Gibaud:2016, Perge:2014}.

The microstructural properties of the carbon black dispersion are investigated using TRUSAXS measurements carried out on the ID02 High Brilliance beamline at the European Synchrotron Radiation Facility (ESRF, Grenoble, France) \cite{Narayanan:2018}. The incident X-ray beam of wavelength 0.1~nm is collimated to a vertical size of 80~$\mu$m and a horizontal size of 150~$\mu$m. A sample-to-detector distance of 31~m is used and provides access to scattering magnitudes $q$ of the scattering wave vector from 0.001 to 0.155~nm$^{-1}$. This corresponds to length scales $2\pi/q$ ranging from about 40~nm to 6.3~$\mu$m. A flow cell is used to measure the intensity scattered by the mineral oil background and the carbon black dispersion at concentration $c_w$. The background scattering from the mineral oil is systematically subtracted to the two-dimensional scattering patterns of the carbon black gel. \tg{As shown in Fig.~\ref{fig:I2D} in the Appendix, the resulting scattering intensity always remains isotropic in the range of flow and ultrasound intensities tested. Therefore,} only radially-average spectra $I(q)$ are presented in the following.

Carbon black particles are fractal aggregates made of fused primary particles of typical diameter 20~nm \cite{Ehrburger1990,Richards:2017}. Using TRUSAXS, we measure $I(q)$ in carbon black dispersions at different weight concentrations $c_w$, see Fig.~\ref{fig:saxscb}a. At the lowest concentration $c_w=0.01$\%~w/w, the particles are isolated and the scattered intensity is proportional to the form factor. This form factor is composed of a plateau at low $q$, a cut-off frequency around $0.04$~nm$^{-1}$ followed by a power-law decay of exponent $-2.9$, which is typical of carbon black particles~\cite{Richards:2017}. Following Richards et al.~\cite{Richards:2017}, $I$ is fitted using a mass fractal model~\cite{sorensen1992} with a Schulz probability distribution of the particle radius as shown in the inset of Fig.~\ref{fig:saxscb}a. This fit yields that a carbon black particle is fractal with a fractal dimension of 2.2 and an average radius of gyration of 60~nm with a 20\% polydispersity. The polydispersity smears out the 2.2 fractal dimension and leads to a power-law decay with exponent $-2.9$. At higher concentrations, carbon black particles aggregate, which leads to an increase of the forward scattering. At $c_w=6$\%~w/w, the dispersion is a gel as shown by rheological measurements in Fig~\ref{fig:caract_cb}. We may distinguish three regimes in the corresponding scattering spectrum $I(q)$ plotted in red in Fig.~\ref{fig:saxscb}a. For $q>0.04$~nm$^{-1}$, $I$ is due to the particle scattering, i.e., it corresponds to the form factor with the characteristic power-law exponent of $-2.9$ as checked in Fig.~\ref{fig:saxscb}c. For $q<0.0025$~nm$^{-1}$, the structure corresponds to a fractal assembly of clusters of carbon black particles down to about 2.4~$\mu$m with a fractal dimension $d_{fb}=2.75$ as inferred from the power-law behavior $I(q) \sim q^{-d_{fb}}$ at low $q$ (see also Fig.~\ref{fig:saxscb}b). At intermediate $q$, \tg{the spectrum $I$ is described in detail in Ref.~\cite{Richards:2017, gibaud2019}} and corresponds to the overlap and interference from the two limiting structures at low and high $q$. 

\subsection{Rheology coupled to high-power ultrasound}

 \begin{figure*}[htb!]
 	\centering
 	\includegraphics[width=0.95\textwidth]{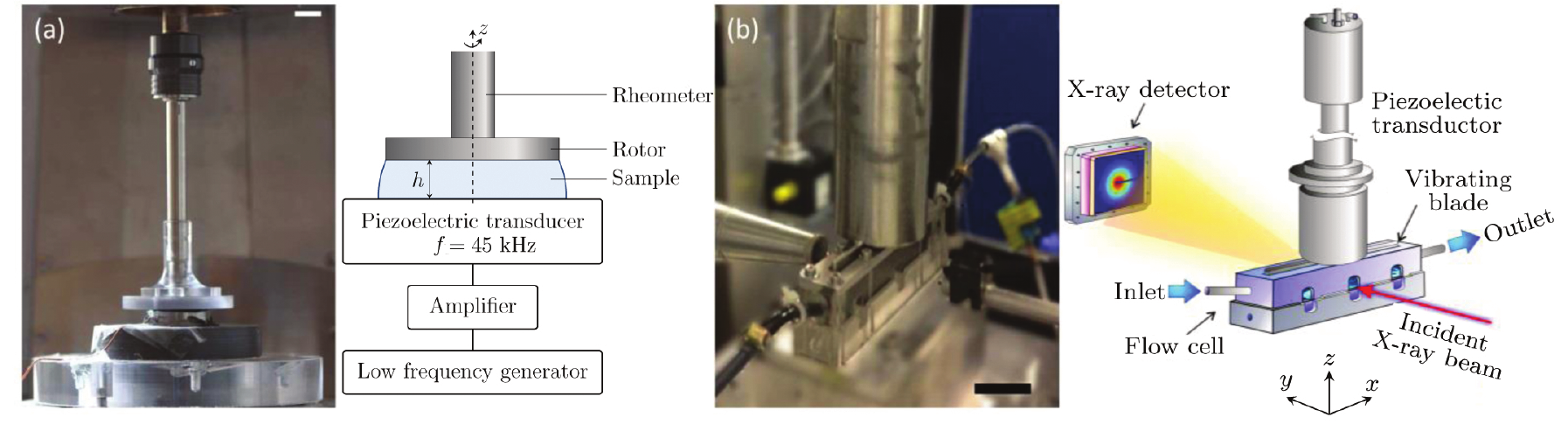}
         \caption{{\bf Rheology and TRUSAXS coupled to high-power ultrasound}. (a)~Rheology coupled to high-power ultrasound: picture (scale bar: $1~$cm) and schematic of the setup. (b)~TRUSAXS coupled to high-power ultrasound: picture (scale bar: $5~$cm) and schematic of the setup.
         }
     \label{fig:setup}
 \end{figure*}

The mechanical effects of ultrasound on colloidal gels are investigated thanks to a rotational rheometer (Anton Paar MCR 301) equipped with an upper rotating geometry which is either a plate of diameter $D=50$~mm made of sandblasted Plexiglas or a cone of diameter 25~mm made of steel, Fig.~\ref{fig:setup}a. The bottom plate is constituted of a piezoelectric transducer working at frequency $f_{US}=45$~kHz (Sofranel, radius $R=14.5$~mm). The piezoelectric transducer is fed with an oscillating voltage of amplitude up to 90~V through a broadband power amplifier (Amplifier Research 75A250A) driven by a function generator (Agilent 33522A). A thermocouple (National Instruments USB-TC01) monitors the temperature at the surface of the transducer. The gap width is set to $h=1$~mm for all experiments. The displacement amplitude $a_{US}$ of the transducer surface along the vertical direction is calibrated using a laser vibrometer (Polytec OVF-505). The highest achievable amplitude is $a_{US}\simeq 0.33$~$\mu$m, much smaller than the gap width $h$. This amplitude corresponds to an acoustic pressure $P=2\pi \rho c\,f_{US}\,  a_{US}\simeq 110$~kPa and to an acoustic power $\mathcal{P}=P^2/\rho c \simeq 1$~W\,cm$^{-2}$, where $\rho=838$~kg\,m$^{-3}$ is the mineral oil density and $c=1.42\cdot 10^3$~m.s$^{-1}$ is the speed of sound in the mineral oil.

 \begin{figure}[hbt!]
 	\centering
 	\includegraphics[width=1\columnwidth]{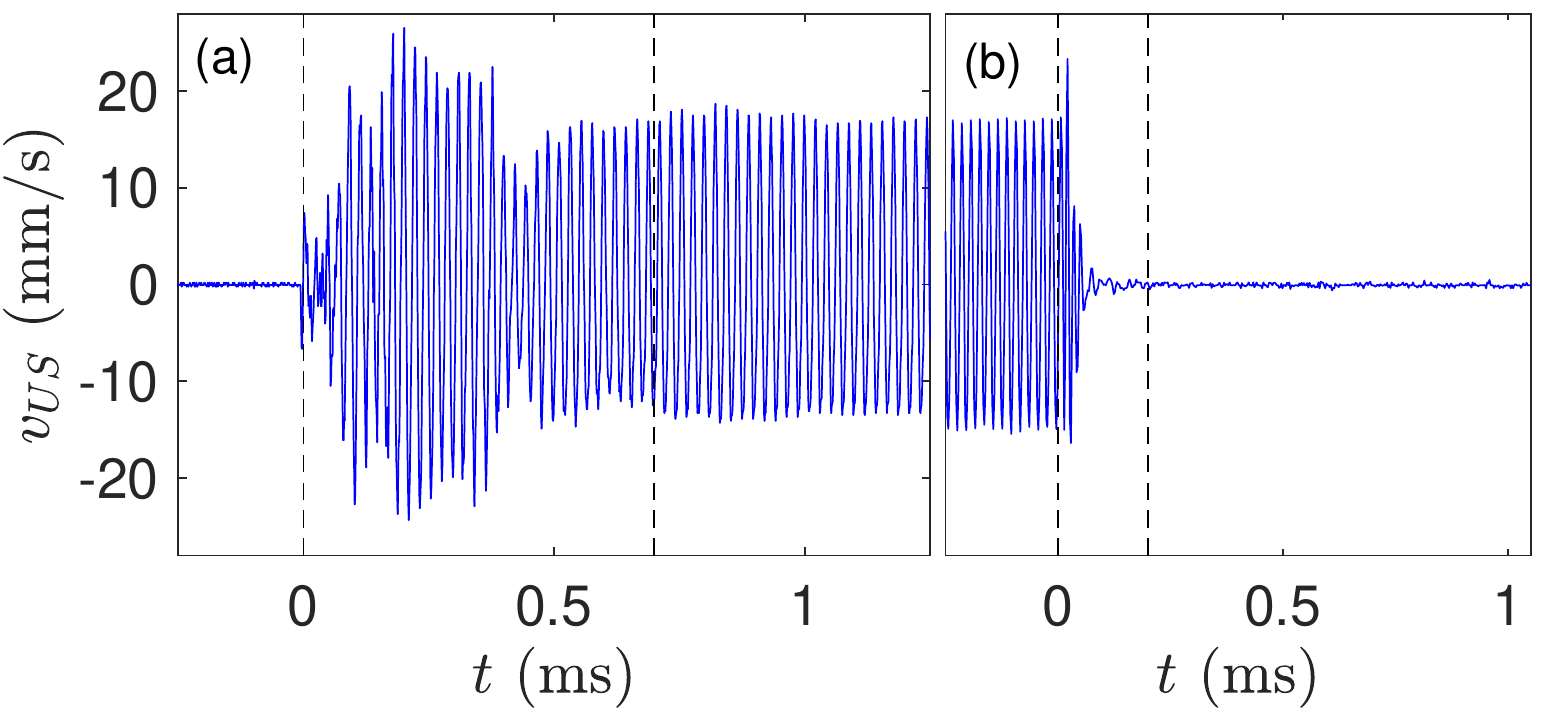}
         \caption{{\bf Temporal response of the piezoelectric transducer used in the rheo-ultrasonic setup.} Velocity $v_{US}$ of the transducer surface measured with a laser vibrometer as a function of time $t$ as ultrasound is (a)~turned on at $t=0$ with an input voltage of frequency $f_{US}=45$~kHz and amplitude 75~V, and (b)~turned off at $t=0$.
         }
     \label{fig:transductor}
 \end{figure}

\tg{The vertical oscillatory displacement of the piezo-electric transducer surface with amplitude $a_{US}$ generates a squeeze flow across the parallel-plate geometry characterized by a strain amplitude $\gamma_{US}=Ra_{US}/2h^2$. At the largest ultrasound amplitudes, $\gamma_{US}$ reaches about 0.24~\%, not far from $\gamma_{NL}\simeq 0.8\%$, so that significant effects are expected on the structure and rheology of the present carbon black gel, as reported in Ref.~\cite{gibaud2019}.}
We measured that, after one minute under ultrasound, the temperature of the transducer surface (and thus of the sample) increases by less than $\Delta T=0.1^\circ$C for $P<60$~kPa, by about 0.2 to 1$^\circ$C for $60<P<100$~kPa and by up to 4$^\circ$C for $P>100$~kPa. As checked in Ref.~\cite{gibaud2019}, this increase in temperature cannot account for the softening of the gel observed when ultrasound is turned on,
\tg{\seb{especially because} an increase of temperature leads to an increase of the carbon black elastic modulus~\cite{won2005}.}
Finally, as shown in Fig.~\ref{fig:transductor}, the amplitude of the piezoelectric transducer displacement reaches a steady state within a few hundreds of microseconds, which allows us to safely probe the gel response to high-power ultrasound on time scales larger than 1~ms, \tg{i.e., seconds}.

\subsection{TRUSAXS coupled to high-power ultrasound}

The effects of ultrasound on the structure of the 6\%~w/w carbon black gel structure are investigated thanks to the TRUSAXS setup under ultrasound described in full details in Ref.~\cite{Jin:2014a} and sketched in Fig.~\ref{fig:setup}b. Briefly, a vertical titanium vibrating blade of width 2~mm connected to a piezoelectric transducer working at $f_{US}=20$~kHz (SODEVA TDS, France) is immersed in a channel of width $d=4$~mm, depth $l=8$~mm and length 100~mm. Using a syringe pump (Harvard Apparatus PHD 4400), the gel is made to flow at a flow rate $Q$ in the TRUSAXS-ultrasound cell. Assuming a Poiseuille flow, the mean shear rate is given by $\bar{\dot{\gamma}}=\frac{3Q}{ld^2}$ and can be tuned from 0 to 10~s$^{-1}$. \tg{Note that this mean shear rate is simply indicative as wall slip and pluglike flows are likely for such low values, as discussed below in Section~\ref{sec:flow}.} A pair of pressure sensors measures the pressure difference $\Delta\Pi$ between the inlet and the outlet of the TRUSAXS-ultrasound flow cell. Ultrasound is transmitted to the gel via the blade vibrating with an amplitude up to $a_{US}=1.6$~$\mu$m corresponding to an acoustic pressure $P\simeq 240$~kPa and an acoustic power $\mathcal{P} \simeq 4.7$~W\,cm$^{-2}$ at the blade surface \cite{Jin:2014a}. TRUSAXS measurements are performed on the carbon black gel at a distance of 1~mm from the vibrating blade through a small window of thickness 0.3~mm machined in the polycarbonate channel wall in the middle of the length of the channel.

\tg{Note that the levels of acoustic power involved in this setup, and \textit{a fortiori} in the previous, less powerful rheo-ultrasonic setup, are not high enough to trigger acoustic cavitation of the suspending mineral oil considered here, i.e., to generate bubbles of oil vapor. It is also very unlikely that ``cavitation'' occurs in the looser sense of bubble formation from gas dissolved in the mineral oil (if any), as this should have a clear signature in the SAXS scattering signal characterized by a tremendous increase of the contrast and therefore of the scattering intensity. Furthermore, previous studies using the same setup on various other systems did not report any significant cavitation activity from the scattering spectra either \cite{Jin:2014a,Jin:2014b,Jin:2015}.}


 \begin{figure*}[htb!]
 	\centering
 	\includegraphics[width=0.85\textwidth]{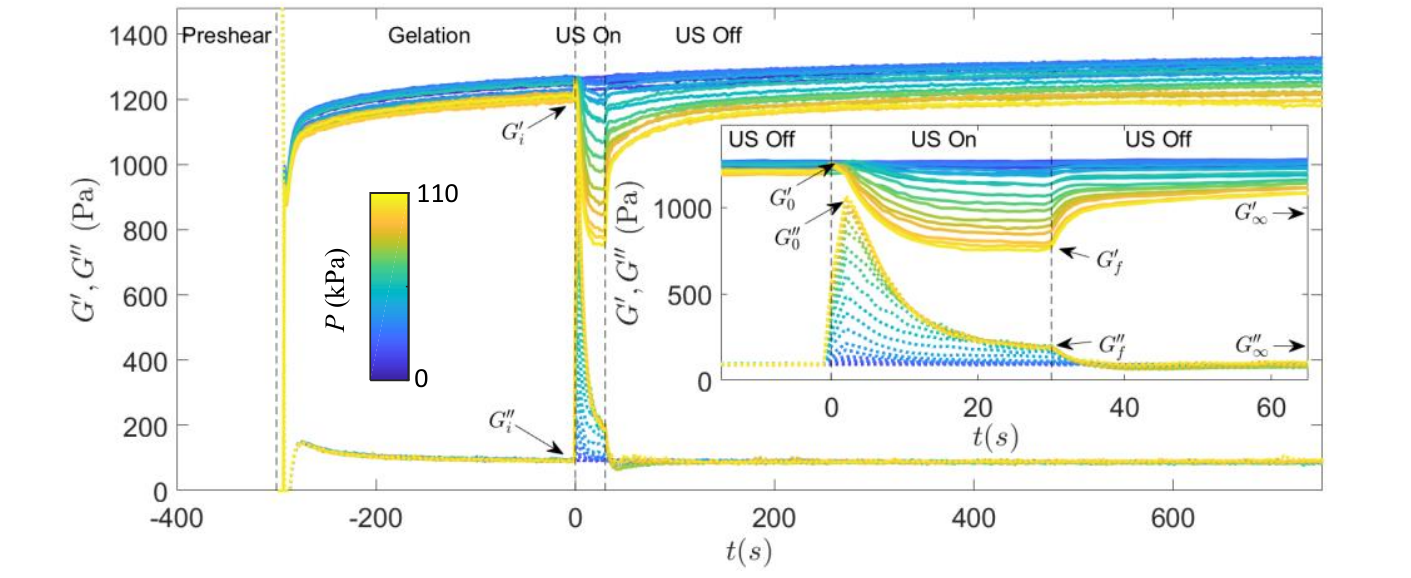}
         \caption{{\bf Effect of high-power ultrasound on the viscoelastic properties of the carbon black gel.} After preshear is stopped at $t = -300$~s, $G'$ is measured as a function of time in the linear regime with a strain amplitude $\gamma=0.06\%$ and frequency $f=1$Hz. After 300~s, the gel reaches a value of $G_i'\simeq 1200 $~Pa and $G_i''\simeq 100$~Pa which we consider as a reference before ultrasound is turned on. Ultrasound is turned on at $t = 0$~s and turned off at $t$ = 30 s. The color codes for the acoustic pressure $P$ from 0 (blue) to $110$~kPa (yellow) \tg{using a linear scale as indicated by the color bar}. Inset: zoom during the time ultrasound is applied. We take the following notations: $G'_0=G'(t=1~$s$)$, $G'_f=G'(t=30~$s$)$, $G'_{\infty}=G'(t=700$~s$)$, $G''_0=G''(t=3~$s$)$, $G''_f=G''(t=30~$s$)$ and $G''_{\infty}=G''(t=700$~s$)$. These experiments are performed in the parallel-plate geometry. \tg{As checked in Fig.~\ref{fig:ss_us}, measurements of $G'$ and $G''$ remain in the linear viscoelastic regime even when ultrasound is turned on.}
         }
     \label{fig:G_HPU}
 \end{figure*}

\section{Carbon black gels at rest under high-power ultrasound}
\label{s:rest}
\subsection{Effect of high-power ultrasound on the mechanical properties}
 \begin{figure}[hbt!]
 	\centering
 	\includegraphics[width=.96\columnwidth]{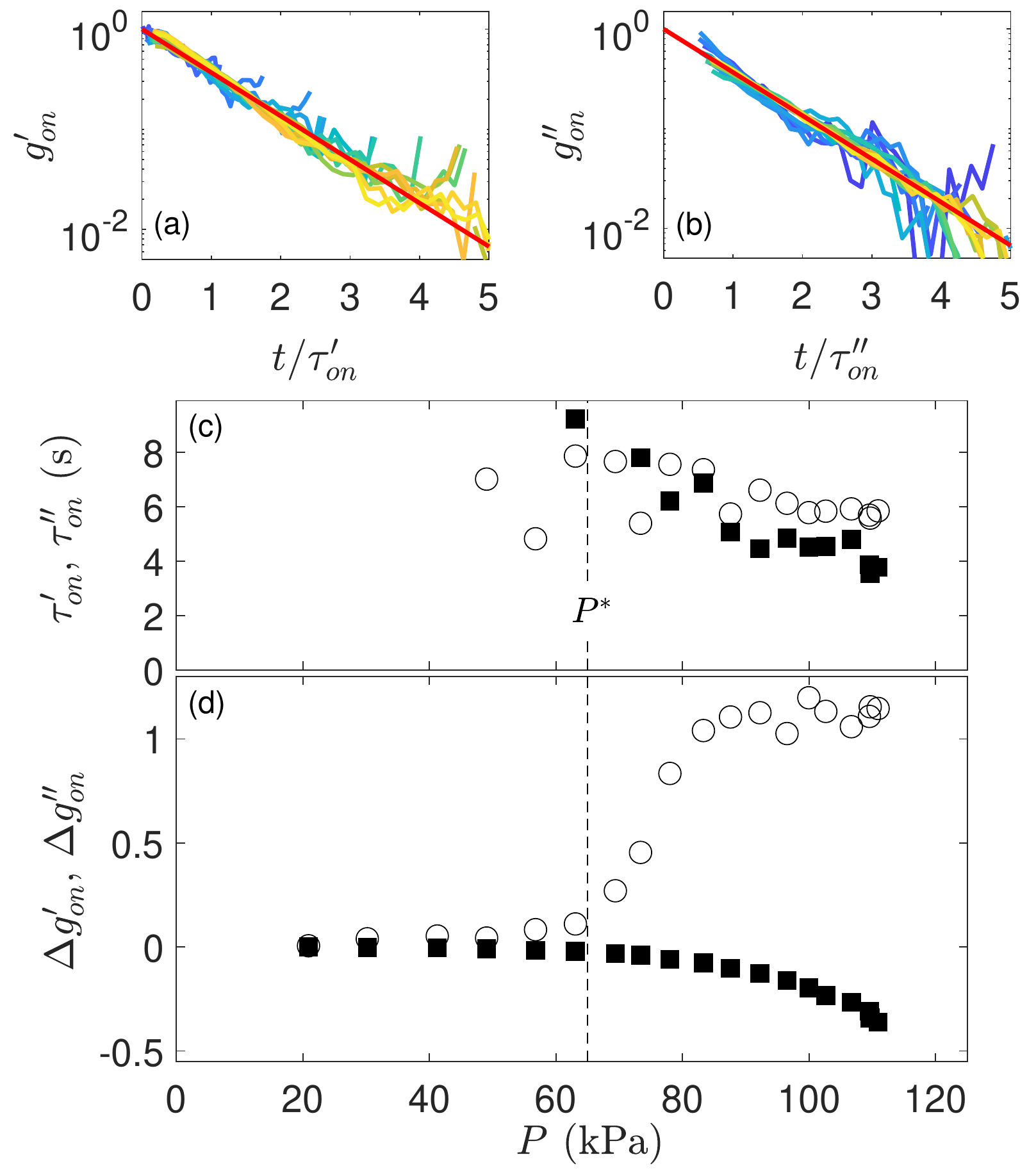}
         \caption{{\bf Softening of the carbon black gel during exposure to high-power ultrasound.} (a)~Normalized value of $\widetilde{G}'$, $g'_{on}=(\widetilde{G}'-\widetilde{G}'_f)/(\widetilde{G}'_0-\widetilde{G}'_f)$, as a function of the normalized time $t/\tau'_{on}$, where $\tau'_{on}$ is the characteristic exponential decay time of $g'_{on}$. (b)~Normalized value of $G''$, $g''_{on}=(\widetilde{G}''-\widetilde{G}''_f)/(\widetilde{G}''_0-\widetilde{G}''_f)$, as a function of the normalized time $t/\tau''_{on}$, where $\tau''_{on}$ is the characteristic exponential decay time of $g''_{on}$. The red curves show the exponential function $y=\exp(-x)$ and \tg{the color scale from $P=0$ (blue) to $110~$kPa (yellow) is the same} as in Fig.~\ref{fig:G_HPU}. (c)~Characteristic decay times $\tau'_{on}$ (full squares) and $\tau''_{on}$ (empty circles) as a function of the acoustic pressure $P$. (d)~Relative variations $\Delta g'_{on}=(\widetilde{G}'_f-G'_{i})/G'_{i}$ (full squares) and $\Delta g''_{on}=(\widetilde{G}''_f-G''_i)/G''_{i}$ (empty circles) as a function of $P$. $P^*$ indicates the pressure value for which ultrasound starts to have a significant effect on the viscoelastic moduli.
         }
     \label{fig:softening}
 \end{figure}
 \begin{figure}[hbt!]
 	\centering
 	\includegraphics[width=.95\columnwidth]{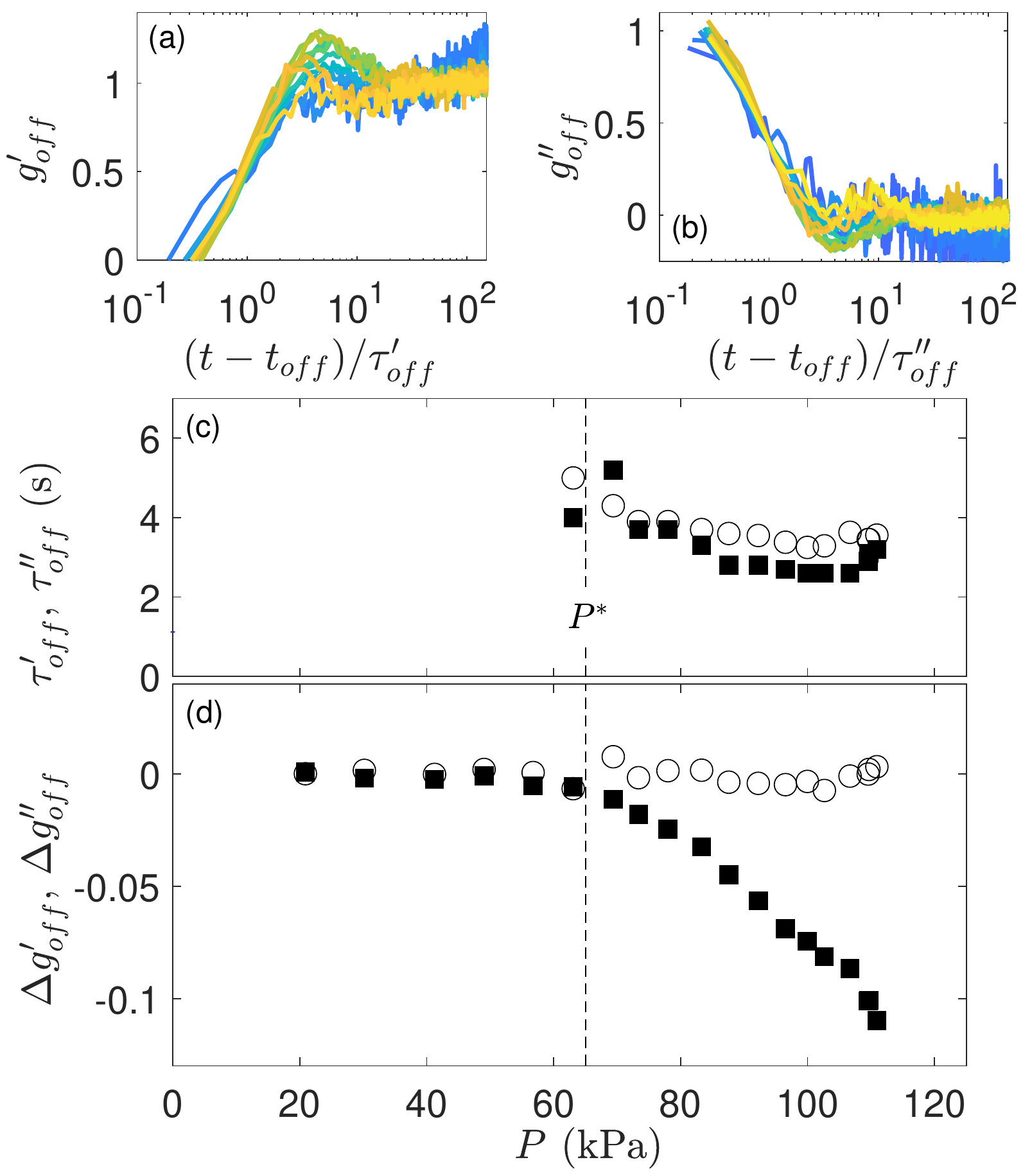}
         \caption{{\bf Recovery of the carbon black gel after exposure  to high-power ultrasound.} (a)~Normalized value of $\widetilde{G}'$, $g'_{off}=(\widetilde{G}'-\widetilde{G}'_f)/(\widetilde{G}'_f-\widetilde{G}'_{\infty})$, as a function of the normalized time $t/\tau'_{off}$, where $\tau'_{off}$ is the characteristic time of $g'_{off}$. (b)~Normalized value of $G''$, $g''_{off}=(\widetilde{G}''-\widetilde{G}''_f)/(\widetilde{G}''_f-\widetilde{G}''_{\infty})$, as a function of the normalized time $t/\tau''_{off}$, where $\tau''_{off}$ is the characteristic time of $g''_{on}$. $\tau'_{off}$ and $\tau''_{off}$ respectively correspond to times at which $g'_{off}=0.5$ and $g''_{off}=0.5$. \tg{The color scale for $P$ is the same as in Fig.~\ref{fig:G_HPU}.} (c)~Characteristic times $\tau'_{off}$ (full squares) and $\tau''_{off}$ (empty circles) as a function of the acoustic pressure $P$. (d)~Relative variations $\Delta g'_{off}=(\widetilde{G}'_{\infty}-G'_i)/G'_{i}$ (full squares) and $\Delta g''_{off}=(\widetilde{G}''_{\infty}-G''_i)/G'_{i}$ (empty circles) as a function of $P$. $P^*$ indicates the pressure value for which ultrasound starts to have a significant effect on the viscoelastic moduli.
         }
     \label{fig:recovery}
 \end{figure}

\subsubsection{\tg{Transient evolution of the viscoelastic moduli}}

\tg{As already reported in Ref.~\cite{gibaud2019},} carbon black gels are highly sensitive to high-power ultrasound. Such sensitivity is \tg{further investigated here} thanks to the rheo-ultrasonic setup. After the preshear protocol, the rheometer monitors the gel viscoelastic properties at rest through small-amplitude oscillatory strain (SAOS) of amplitude 0.06\% and frequency 1~Hz. During the first 300~s, the gels builds up and reaches a value of $G'_i\simeq 1200 $~Pa and $G''_i\simeq 100$~Pa which we will consider as a reference when measuring the effects of ultrasound. Ultrasound is then applied for 30~s and turned off for the rest of the experiment. As shown in Fig.~\ref{fig:G_HPU}, when ultrasound is turned on at $t=0$~s with sufficient power, we observe a softening effect: the elastic modulus $G'$ starts to decrease from a value \tg{$G'_0\simeq G'_i$} at $t\simeq 1$~s down to a minimum value $G'_f$ at $t=30$~s. Meanwhile, the viscous modulus $G''$ shows a strong overshoot and peaks at a value $G''_0$ for $t \simeq 3$~s before decreasing down to \tg{a steady-state value $G''_f$ under ultrasound}. After ultrasound is turned off, a recovery process takes place \tg{within} the gel \tg{and both moduli $G'$ and $G''$ relax to new steady-state values} defined by $G'_{\infty}$ and $G''_{\infty}$ at $t=700$~s. 

\tg{The same preshear and rheo-ultrasonic protocol is repeated in Fig.~\ref{fig:G_HPU} for various acoustic pressures $P$ (from blue corresponding to low pressure levels to yellow for large pressure levels). Qualitatively, it is clear that the decrease of the elastic modulus and the overshoot in the loss modulus are both strongly amplified when $P$ is increased (see also inset in Fig.~\ref{fig:G_HPU}). Moreover, above some critical pressure, it appears that the storage modulus does not fully recover its initial value, i.e., $G'$ does not increase back from its steady-state value $G'_f$ under ultrasound to its initial value  $G_i'$ at least on time scales of 700~s. However, the loss modulus always relaxes rapidly from $G''_f$ down to its initial value $G''_{\infty}=G_i''$. Note that in order to correctly interpret $G'$ and $G''$ measurements, it should be checked that the amplitude $\gamma=0.06\%$ of the oscillatory strain remains in the linear viscoelastic regime of the gel even when ultrasound is turned on. This is confirmed below in Fig.~\ref{fig:ss_us}, which reports strain sweep measurements performed once a steady state is reached under ultrasound.}

\tg{Let us now quantify more precisely the effects of ultrasound on the gel viscoelastic moduli as a function of $P$, first by focusing on the transient toward steady-state under ultrasound.} In order to distinguish between the effects of ultrasound and the weak yet noticeable spontaneous aging of the gel, we first remove aging from $G'$ and $G''$ by fitting $G'$ and $G''$ with logarithmic behaviors for $-250~\text{s}<t<0$ (see red lines in Fig.~\ref{fig:caract_cb}c) and by hereafter working on the corrected moduli $\widetilde{G}'(t)=G'(t)G'_i/(a'\log(t+300)+b')$ and $\widetilde{G}''(t)=G''(t)G''_i/(-a''\log(t+300)+b'')$, where $(a', a'')$ and $(b', b'')$ are respectively the slopes and intercepts of the logarithmic fits of $G'$ and $G''$. As seen in Fig.~\ref{fig:softening}, we can identify two regimes depending on $P$. Below a critical pressure $P^*\simeq 65$~kPa, ultrasound has almost no effect on the viscoelastic moduli of the gel. Above $P^*$, however, the influence of $P$ on $\widetilde{G}'$  and $\widetilde{G}''$ can be mapped onto a single, exponential master curve as shown in Fig.~\ref{fig:softening}a-b: $g'_{on}=(\widetilde{G}'-\widetilde{G}'_f)/(\widetilde{G}'_0-\widetilde{G}'_f)=\exp(-t/\tau'_{on})$ and $g''_{on}=(\widetilde{G}''-\widetilde{G}''_f)/(\widetilde{G}''_0-\widetilde{G}''_f)=\exp(-t/\tau''_{on})$. The characteristic times $\tau'_{on}$ and $\tau''_{on}$ of this exponential decay during the softening effect are close to each other and decrease with increasing $P$ by a factor of about 2, see Fig.~\ref{fig:softening}c. Moreover, the effect of ultrasound on the elastic modulus with respect to $G'_i$ can be quantified by $\Delta g'_{on}=(\widetilde{G}'_f-G'_i)/G'_i$. As shown in Fig.~\ref{fig:softening}d with full squares, increasing the acoustic pressure leads to a larger drop in the gel elasticity,  reaching about 35\% at the highest pressure achievable with our setup. The long-term effect of ultrasound on the loss modulus with respect to $G''_i$ can be similarly quantified by $\Delta g''_{on}=(\widetilde{G}''_f-G''_i)/G''_i$, which disregards the transient overshoot in $G''$. As shown by the open circles in Fig.~\ref{fig:softening}d, increasing $P$ leads to an increase of the gel loss modulus by about 130\% at the highest ultrasonic pressure. 

We now focus on quantifying the gel recovery once ultrasound with acoustic pressure $P$ is turned off. We define quantities $g'_{off}$, $g''_{off}$, $\Delta g'_{off}$ and $\Delta g''_{off}$ similar to those introduced for the previous phase but based on the age-corrected steady-state values $ \widetilde{G}'_{\infty}$ and $\widetilde{G}''_{\infty}$ of the viscoelastic moduli after exposure to ultrasound. Here again, when properly rescaled, the recovery of both $\widetilde{G}'$ and $\widetilde{G}''$ can be mapped onto a master curve, as shown in Fig.~\ref{fig:recovery}a-b. Within a duration of about 4~s, $\widetilde{G}'$ and $\widetilde{G}''$ respectively follow symmetric increasing and decreasing trends before both moduli reach a plateau. This characteristic relaxation time of about 4~s does not depend significantly on $P$, as seen in Fig.~\ref{fig:recovery}c. Remarkably, the recovery process is always complete for $G''$, i.e., $\Delta g''_{off}\simeq 0$ for all values of $P$, while it is only partial for $G'$ whenever ultrasound has a significant effect on the viscoelastic moduli: as $P$ increases above $P^*$, the recovery of $G'$ is all the more incomplete, see Fig.~\ref{fig:recovery}d. To sum up, increasing the acoustic pressure does not affect the recovery time but leads to incomplete recovery above $P^*$, where the gel remains less elastic than prior to sonication.

\subsubsection{\tg{Steady-state viscoelastic properties under ultrasound}}

\begin{figure*}[htb!]
 	\centering
 	\includegraphics[width=0.9\textwidth]{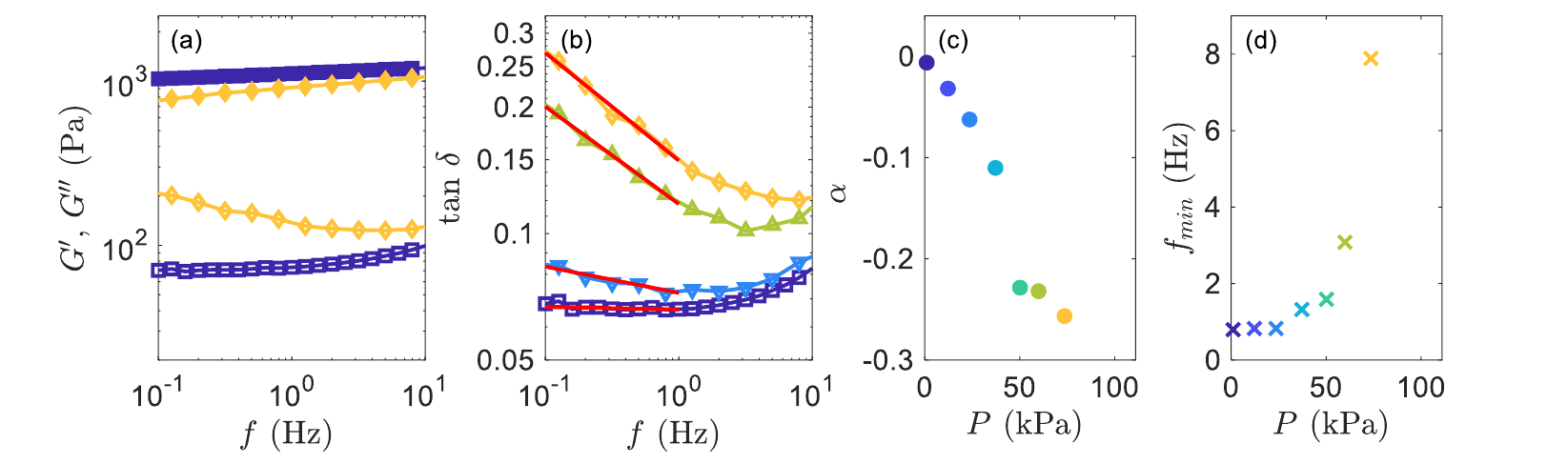}
         \caption{\tg{{\bf Influence of high-power ultrasound on the viscoelastic spectrum}. (a)~Storage modulus $G'$ (full symbols) and loss modulus $G''$ (empty symbols) of the carbon black gel as function of frequency $f$ in the absence of ultrasound, i.e. for $P=0$ ($\square$) and in the presence of ultrasound with acoustic pressure $P=73$~kPa ($\diamond$) measured in the parallel-plate geometry with a strain amplitude $\gamma=0.06\%$ at a rate of 2 cycles per point, leading to a total duration of 381~s for the whole frequency sweep. (b)~Loss tangent $\tan\delta=G''/G'$ as a function of $f$ for $P=0$ ($\square$), 25 ($\triangledown$), 60 ($\triangle$) and 73~kPa ($\diamond$). Red lines are power-law fits between for $f=0.1$--1~Hz such that $\tan\delta\sim f^{\alpha}$. (c)~Power-law exponent $\alpha$ as a function of $P$. Above $P=73$~kPa, ultrasound  strongly heats the gel ($\Delta T> 10^{\circ}$C over the 381~s duration of the frequency sweep) and the corresponding data were therefore discarded. (d)~Evolution of $f_{min}$ as a function of $P$. $f_{min}$ corresponds to the frequency at which $\tan\delta$ is minimum.}
         }
     \label{fig:fs_us}
 \end{figure*}

\tg{In order to investigate the viscoelastic spectrum of the gel under high-power ultrasound, we carry out a frequency sweep 20~s after ultrasound is turned on, i.e., once the gel viscoelastic moduli have reached a steady state under ultrasound. Figure~\ref{fig:fs_us}a shows that the frequency-dependence of $G'$ and $G''$ are deeply modified by high-power ultrasound: while both moduli only increase weakly with frequency $f$ in the absence of ultrasound, $G'$ increases more steeply with $f$ while $G''$ decreases at low frequencies and goes through a minimum under ultrasound. This peculiar behavior is best illustrated in Fig.~\ref{fig:fs_us}b which displays the evolution of $\tan\delta=G''/G'$ as function of $f$ for different values of the acoustic pressure $P$. As $P$ increases, dissipation strongly increases at low frequencies and $\tan\delta$ passes through a minimum at a characteristic frequency $f_{min}$. This trend is characterized in Fig.~\ref{fig:fs_us}c through the exponent $\alpha$ inferred from power-law fits of $\tan\delta$ vs $f$ at low frequencies and in Fig.~\ref{fig:fs_us}d with the evolution of $f_{min}$ as a function of $P$. As shown in  Fig.~\ref{fig:fs_us}c, $\alpha$ is negative and \seb{decreases with increasing} $P$. As shown in  Fig.~\ref{fig:fs_us}d, the minimum in $G''$ shifts towards a larger frequency $f_{min}$ as $P$ increases. Note that frequency sweeps could be reliably measured only for $P<73$~kPa due to too strong heating of the sample under larger acoustic pressures applied for a long period of time ($\sim 400$~s).} 

\begin{figure}[hbt!]
 	\centering
 	\includegraphics[width=1\columnwidth]{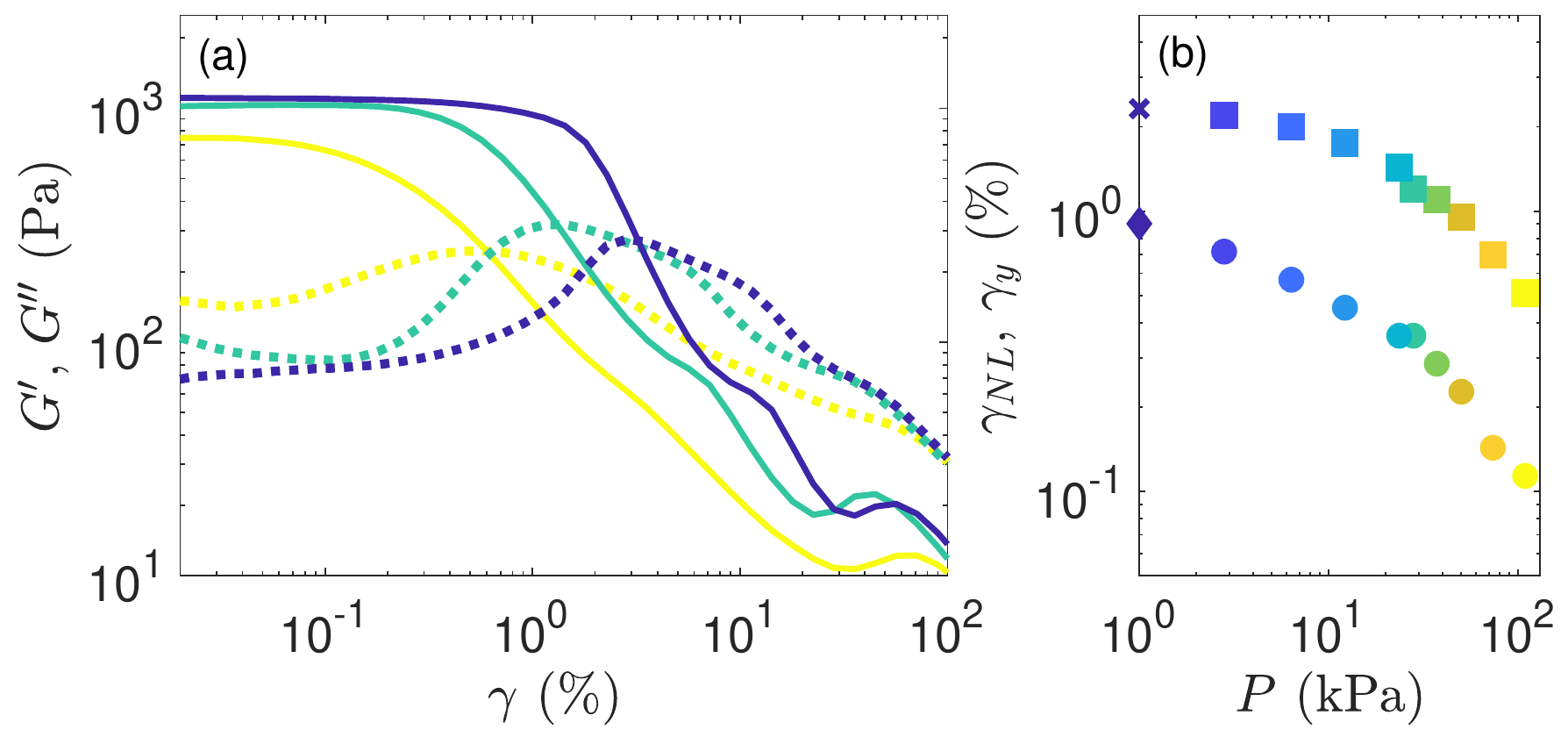}
         \caption{\tg{{\bf Strain sweep experiments under high-power ultrasound.} (a)~Storage modulus $G'$ (solid lines) and loss modulus $G''$ (dotted lines) measured at $f=1$~Hz with a waiting time of 2~s per point as a function of the strain amplitude $\gamma$ for different values of the acoustic pressure: $P=0$ (blue), $P=30$~kPa (cyan) and $P=110$~kPa (yellow). The total duration of the strain sweep experiment is 96~s. As in Fig.~\ref{fig:caract_cb}b, $\gamma_{NL}$ is the lower limit of the non-linear regime and $\gamma_y$ is the yield strain. (b)~Characteristic strains $\gamma_{NL}$ ($\bullet$) and $\gamma_y$ ($\blacksquare$) as a function of $P$. The diamond and cross symbols at $P=1$~kPa correspond respectively to $\gamma_{NL}$ and $\gamma_y$ actually measured at $P=0$~kPa. The color scale for $P$ is the same as in Fig.~\ref{fig:G_HPU}. These experiments are performed in the parallel-plate geometry.}
         }
     \label{fig:ss_us}
 \end{figure}

\tg{Next, in order to further quantify the viscoelastic response under ultrasound in the nonlinear regime, we perform strain sweeps experiments at $f=1$~Hz, starting 20~s after ultrasound is turned on. Figure~\ref{fig:ss_us}a reveals that the evolution of $G'$ and $G''$ as the function of the strain amplitude $\gamma$ is similar to one observed at rest in Fig.~\ref{fig:caract_cb}b. Indeed, the linear regime extends up to $\gamma=\gamma_{NL}$, then $G'$ decreases and crosses $G''$ at the yield strain $\gamma=\gamma_{y}$, while $G''$ goes through a broad maximum around $\gamma_{y}$. Above $\gamma_{y}$, $G''>G'$ and the gel is fluidized. However, under the effect of high-power ultrasound, we note that $\gamma_{NL}$ and $\gamma_y$ are shifted towards lower values as compared to the situation without ultrasound. As shown in Fig.~\ref{fig:ss_us}b, those values decreases monotonically by about one order of magnitude with the acoustic pressure $P$, from $\gamma_{NL}\simeq 0.9$\%  and $\gamma_y \simeq 2.4$\% in the absence of ultrasound down to $\gamma_{NL}\simeq 0.11$\% and $\gamma_y \simeq 0.5$\% at the highest acoustic pressure. Finally, we emphasize that the strain amplitude of 0.06\% chosen to perform all viscoelastic measurements in the previous sections remains always below $\gamma_{NL}$ so that $G'$ and $G''$ reported are indeed recorded within the linear regime.}


\tg{To sum up, for $P<P^*\simeq 65$~kPa, ultrasound has a very limited effect on the value of $G'$ and $G''$ measured at $f=1$~Hz. However, it significantly changes both the gel viscoelastic spectrum at low frequencies and the \seb{extent} of its linear regime. For $P>P^*$, increasing the acoustic pressure leads to a complex viscoelastic transient regime of a few seconds characterized by a strong peak in $G''$, and by exponential decays of $G'$ and $G''$. The gel does not fully recover its initial elasticity at the largest acoustic pressures.}

\subsection{Effect of high-power ultrasound on the structure}
\label{s:structure}

\tg{In order to interpret the above rheological measurements, it is necessary to investigate how high-power ultrasound affects the bulk structure of the carbon black gel. To this aim, we turn to} TRUSAXS experiments coupled to high-power ultrasound. The feed channel of the flow cell is filled with the carbon black gel at 6\% wt. thanks to a syringe pump. The gel is then presheared by pumping a volume of 10~mL at a flow rate $Q=20$~mL min$^{-1}$ in both directions, which corresponds to a mean shear rate $\bar{\dot{\gamma}}\simeq\pm 8$~s$^{-1}$ applied for 30~s in each direction. \tg{Since this preshear rate lies significantly below that used in rheological experiments, the initial microstructure of the gel in the TRUSAXS experiments may differ from that in the rheo-ultrasonic setup \cite{Ovarlez:2013,Helal:2016}.} After a rest time of 180~s, ultrasound is applied for 30~s with a pressure amplitude ranging from $P=20$ to 240~kPa. TRUSAXS spectra are recorded every second during 100~s starting 20~s prior to application of ultrasound. Thus, these measurements provide insight into the microstructure under ultrasound for acoustic pressures similar to and up to twice those of previous rheo-ultrasonic experiments.

 \begin{figure}[hbt!]
 	\centering
 	\includegraphics[width=0.9\columnwidth]{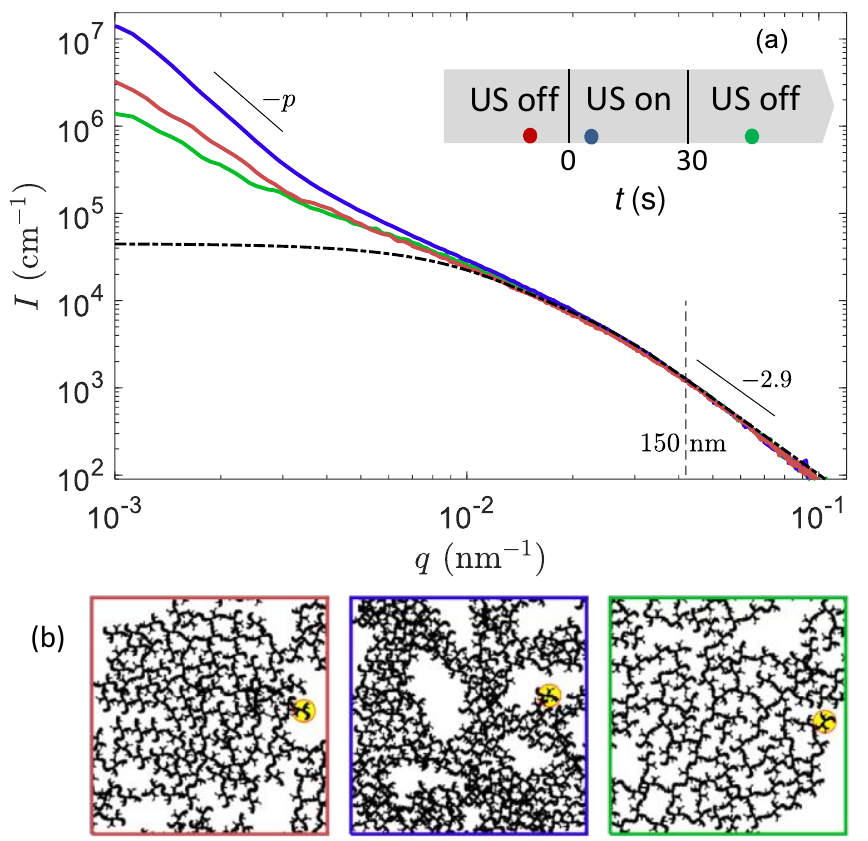}
         \caption{{\bf Structural measurements under ultrasound.} TRUSAXS intensity spectra $I(q)$ recorded in a $6\%$~w/w carbon black gel at rest and exposed to ultrasound with frequency $20~$kHz for an acoustic pressure $P=240$~kPa. Ultrasound is turned on at $t=0$ and switched off at $t=30$~s. Comparison between the scattered intensity $I$ obtained from TRUSAXS  as a function of the wave number $q$ of the gel at rest (red, $t<0$~s), under ultrasonic vibrations (blue, 0~s$<t<30$~s) and during recovery (green, $t>30$~s). The black dash-dotted curve is the form factor reported in blue in Fig.~\ref{fig:saxscb}a. $p$ is the power law exponent defined by $I(q)\sim q^{-p}$ at low $q$. (b) Sketches of the gel microstructure \tg{inferred from TRUSAXS measurements}: before (top), during (middle) and after application of high-power ultrasound (bottom) as inferred from TRUSAXS data. The yellow circle highlights a primary aggregate of diameter $2r_g=120~$nm.}
     \label{fig:saxs_rest_hpu}
 \end{figure}

\begin{figure}
 	\centering
 	\includegraphics[width=1\columnwidth]{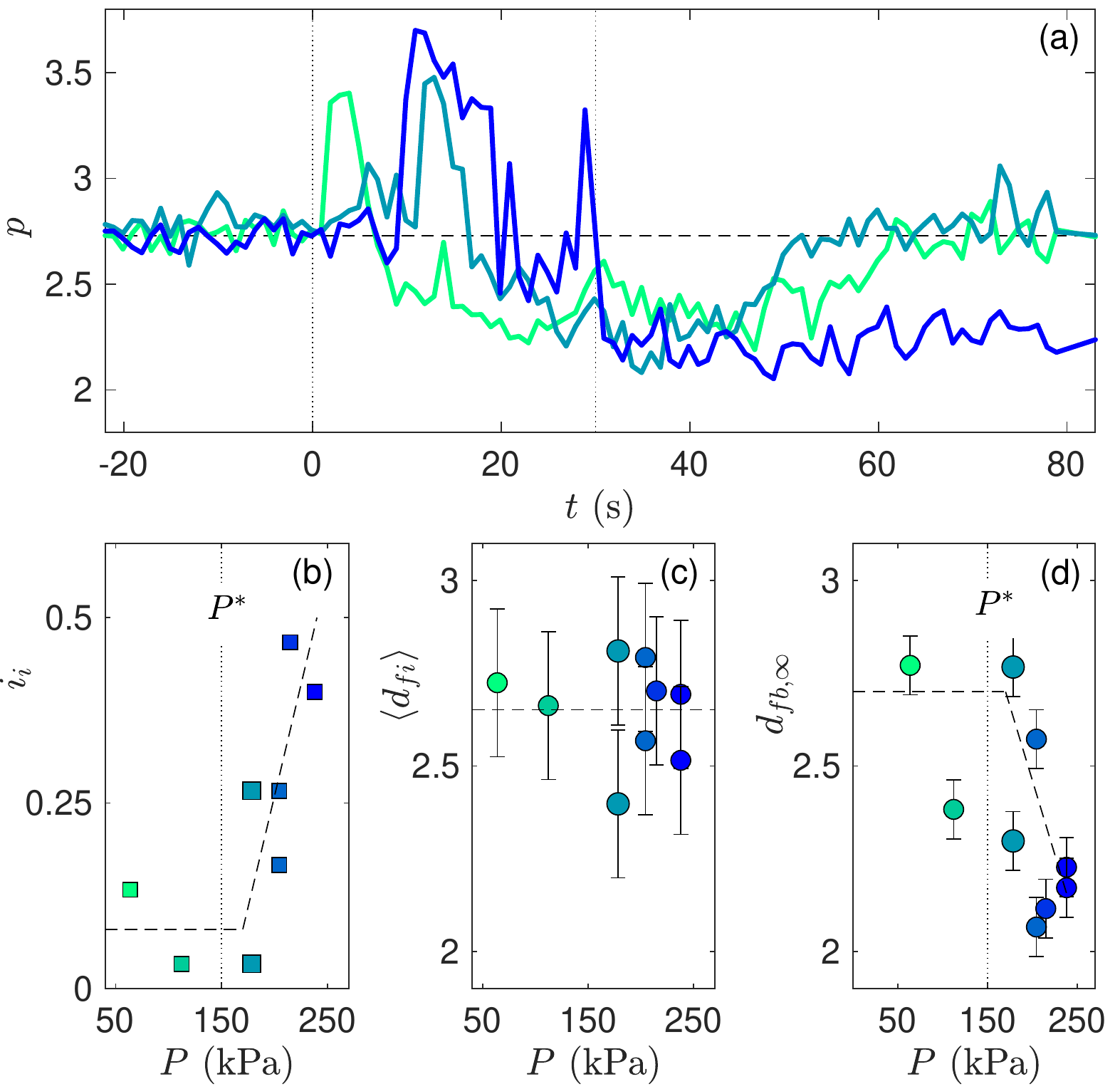}
         \caption{{\bf Influence of the acoustic pressure $P$ on TRUSAXS spectra.} (a)~Exponent $p$ of the best power-law fit $I(q)\sim q^{-p}$ at low $q$ as a function of time. Ultrasound with frequency $20~$kHz is turned on at time $t=0$ and switched off at $t=30$~s. Each curve corresponds to a given acoustic pressure $P=64$ \tg{(green)}, 160 \tg{(blue-green)} and 240~kPa \tg{(blue)}. (b)~Intermittency index $i_i$ as a function of $P$. (c)~Micro-cracks average interfacial fractal dimension $\langle d_{fi} \rangle$ as a function of $P$. (d)~Bulk fractal dimension of the gel $d_{fb,\infty}$ measured 40~s after ultrasound are turned off as a function of $P$.
         }
     \label{fig:saxs_rest_param_hpu}
 \end{figure}

\tg{Based on this TRUSAXS-ultrasound setup, we recently demonstrated in Ref.~\cite{gibaud2019} that ultrasonic vibrations strongly affect the microstructure of carbon black gels. Figure~\ref{fig:saxs_rest_hpu} summarizes our previous findings by displaying three intensity spectra $I(q)$ recorded at rest (red), under high-power ultrasound (blue) and after ultrasound is turned off (green), together with sketches of the corresponding microstructure that can be inferred from $I(q)$.} For scattering wave vector magnitudes $q\gtrsim 2\cdot 10^{-2}$~nm$^{-1}$, the scattering intensity $I(q)$ remains unchanged: ultrasound does not affect the gel structure at length scales smaller than about 300~nm. However for $q<4\cdot 10^{-3}$~nm$^{-1}$, TRUSAXS  measurements indicate that the gel may form cracks. \tg{As discussed in Ref.~\cite{gibaud2019},} such cracks would be filled with oil and the crack interfaces between the gel and the pure oil phase would account for the change of slope $p$ in the scattering intensity $I(q)\sim q^{-p}$, for $q<4\cdot 10^{-3}$~nm$^{-1}$. Indeed, at rest, the gel forms a mass fractal structure~\cite{sorensen1992} and $p$ corresponds to the bulk fractal dimension of the gel, $d_{fb}=p=2.75$, as already reported in Fig.~\ref{fig:saxscb}. Under high-power ultrasound, $p$ may become higher than 3, as shown in Fig.~\ref{fig:saxs_rest_param_hpu}a. In this case, the mass fractal model is no longer valid as $p$ must remains smaller than 3. Such a high value of $p$ is however compatible with scattering from interfaces of fractal dimension $d_{fi}=6-p=2.5$~\cite{beaucage1994}. \tg{Such rough interfaces are likely to be due to oil-filled micro-cracks that nucleate under high-power ultrasound, although a direct visualization of the local microstructure under ultrasound would be necessary to fully confirm this picture}. Those oil-gel interfaces have a higher contrast than the bulk gel and therefore dominate the scattering intensity at low $q$.

\tg{To go beyond the analysis proposed in our previous work \cite{gibaud2019}, we report in Fig.~\ref{fig:saxs_rest_param_hpu} a systematic study of the impact of the acoustic pressure $P$ on the TRUSAXS spectra} during and after application of ultrasound. We first focus on the temporal evolution of the power-law exponent $p(t)$ of $I(q)$ at low $q$ in  Fig.~\ref{fig:saxs_rest_param_hpu}a. At rest, the gel displays a constant value of $p$ equal to $d_{fb}=2.75$. When ultrasound is turned on, we observe intermittent dynamics where $p$ shows bursts that reach values up to 3.7. The effect of ultrasound can be quantified by three parameters: $i_i$ in Fig.~\ref{fig:saxs_rest_param_hpu}b, $\langle d_{fi} \rangle$ in Fig.~\ref{fig:saxs_rest_param_hpu}c and $d_{fb,\infty}$ in Fig.~\ref{fig:saxs_rest_param_hpu}d. 
$i_i$ is the intermittency index defined as $i_i=\Delta t_{p>3}/\Delta t_{US}$ where $\Delta t_{p>3}$ is the total duration over which the spectrum shows an exponent $p$ greater than $3$ and $\Delta t_{US}=30$~s is the duration of exposure to ultrasound.
$\langle d_{fi} \rangle$ is defined as the time average of the interfacial fractal dimension during the bursts, i.e., $\langle d_{fi} \rangle=\langle 6-p(t)\rangle$ for $p>3$. 
$d_{fb,\infty}$ is the mass fractal dimension of the gel 40~s after ultrasound is turned off.

As shown in Fig.~\ref{fig:saxs_rest_param_hpu}c, the fractal dimension of the micro-crack interface remains roughly constant over the entire range of acoustic pressure $P$ covered by our experiments. Its value $\langle d_{fi} \rangle \simeq 2.65 \pm 0.24$ is intermediate between a sharp interface in the Porod limit ($d_{fi}=2$) and a spongy interface ($d_{fi}=3$). However, looking at $i_i$ \tg{in Fig.~\ref{fig:saxs_rest_param_hpu}b}, two regimes can be distinguished below and above $P^* \simeq 150$~kPa: below $P^*$, the occurrence of micro-cracks remains rare, $i_i\simeq 0.05$, \tg{while above $P^*$,} $i_i$ increases sharply, i.e., the occurrence of micro-cracks strongly increases with $P$. \tg{Although experimental scatter is rather larger in Fig.~\ref{fig:saxs_rest_param_hpu}d, the data for the final bulk fractal dimension $d_{fb,\infty}$ are consistent with a full recovery of the initial structure below $P^*$ ($d_{fb,\infty} \simeq 2.75$) and an incomplete recovery above $P^*$, where $d_{fb,\infty}$ shows a decreasing trend from 2.75 down to about 2.1, i.e., the gel structure loosens.}

\subsection{Discussion}

A first observation from the above results is that high-power ultrasound significantly affects the gel properties only above a critical ultrasonic pressure $P^*$. \tg{Such a threshold pressure is most likely linked to the fact that the present material is a soft solid that presents a yield stress. In particular, the strain associated with the acoustic pressure $P^*$ is $\gamma^*_{US}=0.13\%$, which is comparable to $\gamma_{NL}\simeq 0.8\%$, so that significant effects are expected as reported in Ref.~\cite{gibaud2019}.}
We note that the value of $P^*$ is higher in the TRUSAXS setup as compared to the rheology setup. This can be attributed to fact that the shearing geometries, the preshear protocols and the ultrasonic frequencies differ in the two experimental setups. In particular, the vibration field generated 1~mm below the vibrating blade in the TRUSAXS setup most likely differs strongly from the one generated by the vibrating piezoelectric disk in the more confined parallel-plate geometry. Since the values of $a_{US}$ and $P$ reported in our measurements correspond to those at the vibrating surfaces, it is not surprising to find that $P^*$ is larger for the TRUSAXS setup where the blade geometry and the larger sample volume are expected to reduce the efficiency of the vibrations on the colloidal gel.

Focusing on the rheology alone, we observe that both the storage modulus $G'$ and the loss modulus $G''$ are affected by high-power ultrasound for $P>P^*$. Upon turning on ultrasound, a transient regime of a few seconds is observed, where the gel always behaves as a soft solid, i.e., $G'$ remains larger than $G''$. Still, $G'(t)$ displays a monotonic decrease while $G''(t)$ strongly peaks before decreasing in a manner similar to $G'(t)$. This overshoot is reminiscent of the increase of $G''$ by a factor of about 3 observed prior to the yielding transition between $\gamma_{NL}$ and $\gamma_y$ and followed by a decrease of $G''$ for $\gamma>\gamma_y$ (see Fig.~\ref{fig:caract_cb}b).


Combining the results from rheo-ultrasonic experiments with structural measurements under high-power ultrasound allows us to further interpret the evolution of the viscoelastic moduli in light of the formation of micro-cracks revealed by TRUSAXS. Such micro-cracks may be seen as rupture precursors, in the sense of structural rearrangements that locally damage the mechanical properties of the gel but that do not grow large enough to percolate and to fully fluidize the sample~\cite{cipelletti2020, Gibaud2020}. \tg{The intermittency inferred from the scattering by the micro-crack interfaces supports the fact that such micro-cracks open and close in time}. The nucleation and proliferation of these micro-cracks \tg{for $P>P^*$} may explain the very strong transient overshoot in $G''$, by a factor of about 10, upon turning on ultrasound. In steady state, micro-cracks would then lead to an enhanced dissipative contribution, with $G''_f\simeq 2G''_i$ for large acoustic pressures (see Fig.~\ref{fig:softening}d). \tg{This hypothesis is supported by the  viscoelastic spectrum  measurements performed once the system has reached a steady state under ultrasound (Fig.~\ref{fig:fs_us}). Even below $P^*$, we observe a clear evolution of the viscoelastic spectrum, where $\tan\delta=G''/G'$ displays a minimum at a frequency $f_{min}$ that is all the more pronounced that the acoustic pressure is large. The increase of $f_{min}$ with $P$ implies that dissipation mechanisms become preponderant at long time scales. Viscoelastic spectra with a similar minimum in $\tan\delta$ are classically reported in soft ``glassy'' systems \cite{Mason1995elasticity,Mason1995linear,Winter:2013,Zaccone:2014} and attributed to collective rearrangements associated with the $\alpha$-relaxation. In our case, the gel structure is maintained locally. We believe that large length-scale rearrangements are to be attributed to the micro-crack dynamics. Indeed, as $P$ increases, the micro-crack intermittency increases, \seb{consistent} with more large-scale rearrangements, leading to the observed enhancement of dissipation at low frequencies and thus to glass-like viscoelastic features.} 

\tg{Finally, after ultrasound is turned off,} the gel recovery takes a few seconds, consistently with the time scale for gel reformation after a strong preshear (see Fig.~\ref{fig:caract_cb}c). The fact that the elastic modulus does not fully recover its initial value \tg{for $P>P^*$} may be attributed to ultrasound-induced changes in the colloidal flocs whose structure is looser and that can only be reformed as in the initial gel by applying strong enough shear. Finally, the fact that the viscous modulus always recovers its initial value strongly suggests that micro-cracks fully heal once ultrasound is turned off.

\section{Carbon black gels flowing under high-power ultrasound}
\label{sec:flow}

\subsection{Effect of high-power ultrasound on the flow curves}

Our \tg{next} important result is that the flow of carbon black gels is \tg{promoted} by high-power ultrasound. To evidence this effect, we use the rheology experiment coupled to high-power ultrasound. After the preshear protocol, we perform flow curve measurements: the shear rate $\dot\gamma$ is logarithmically swept down from 1000 to 0.01~s$^{-1}$ within $60$~s while the rheometer measures the stress response $\sigma$. Ultrasound is applied during the entire flow curve measurements. This experiment is repeated under different acoustic pressures $P$. 

\begin{figure}
 	\centering
 	\includegraphics[width=0.9\columnwidth]{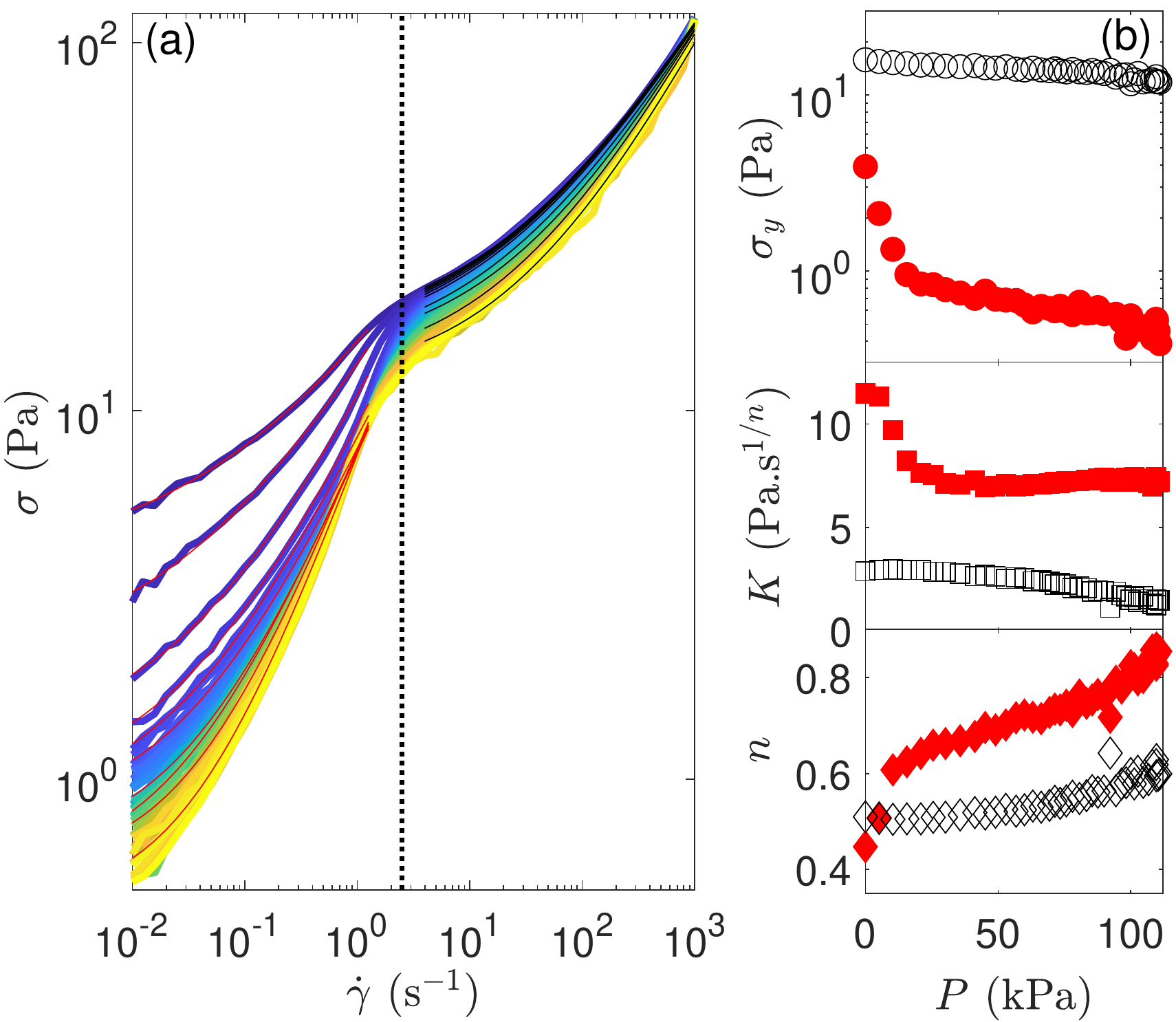}
         \caption{ {\bf Effect of high-power ultrasound on the flow properties of the carbon black gel.} (a)~Flow curves, shear stress $\sigma$ vs shear rate $\dot\gamma$, for different acoustic pressures $P$. \tg{The color scale from $P=0$ (blue) to $110~$kPa (yellow) is the same as in Fig.~\ref{fig:G_HPU}.} Lines are fits with the Herschel-Bulkley models performed at low shear rates (for $\dot\gamma<1$~s$^{-1}$, red lines) and at high shear rates (for $\dot\gamma>5$~s$^{-1}$, black lines). The vertical dashed line at $\dot\gamma^*\simeq 2.5$~s$^{-1}$ indicates the separation between the low and high shear rate branches of the flow curve. (b)~Herschel-Bulkley parameters as a function of the acoustic pressure $P$ (full red symbols for $\dot\gamma<1$~s$^{-1}$ and hollow black symbols for $\dot\gamma>5$~s$^{-1}$): yield stress $\sigma_y$, consistency index $K$ and flow index $n$ from top to bottom. Experiments performed in the cone-and-plate geometry.
         }
     \label{fig:flowcurve_pierre}
 \end{figure}

As shown in Fig.~\ref{fig:flowcurve_pierre}a, the flow curve can be decomposed into two branches depending on the shear rate value with respect to $\dot\gamma^* \simeq 2.5$~s$^{-1}$. This decomposition is motivated by previous studies showing that such flow curves are typically dominated by wall slip and pluglike flows below $\dot\gamma^*$~\cite{Meeker:2004b,divoux2013,Grenard:2014}. \tg{Such a competition between bulk flow and wall slip is also confirmed in Fig.~\ref{fig:FCgap} in the Appendix by investigating the dependence of the flow curves at low shear rates on the gap width in the parallel-plate geometry. As for the flow curves of Fig.~\ref{fig:flowcurve_pierre}a measured in the cone-and-plate geometry, both the high-shear and low-shear regimes can be well fitted by a Herschel Bulkley model,\cite{herschel1926}} $\sigma= \sigma_y+K\dot\gamma^n$, where $\sigma_y$ is the yield stress, $K$ is the consistency index, and $n$ is the flow index with $n=1$ for a Newtonian fluid and $n<1$ for shear-thinning fluids. As shown in Fig.~\ref{fig:flowcurve_pierre}b with black hollow symbols, ultrasound only mildly affects the high shear branch: the yield stress $\sigma_y$ slightly decreases from $15$ to $12$~Pa and the flow index slightly increases from $0.51$ to $0.62$. In contrast, ultrasound has a much stronger effect on the low shear branch. Indeed, as shown in Fig.~\ref{fig:flowcurve_pierre}b  with full red symbols, as $P$  increases, $\sigma_y$ decreases by more than a decade while $n$ almost doubles from 0.45 up to 0.85. In this regime, high-power ultrasound effectively turns the gel from a yield stress solid into a more and more Newtonian fluid: $\sigma_\text{y} \rightarrow 0$ and $n\rightarrow 1$. To sum up, high power-ultrasound strongly \tg{facilitates} the flow of the carbon black gel at low shear rates below $\dot\gamma^*$ while it affects the flow under high shear rates much more mildly.
\label{s:flowcurve}

\subsection{Effect of high-power ultrasound on the structure under flow}
 \begin{figure}
 	\centering
 	\includegraphics[width=0.9\columnwidth]{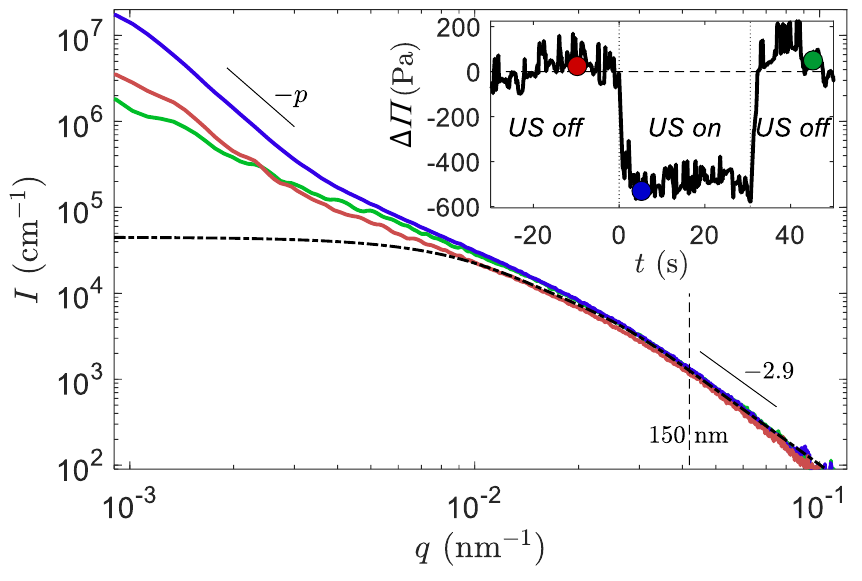}
         \caption{ {\bf Structural measurements under flow and ultrasound.} TRUSAXS intensity spectra $I(q)$ recorded in a flowing $6\%$ w/w carbon black gel exposed to ultrasound with frequency $20~$kHz for an acoustic pressure $P=240$~kPa. A mean shear rate $\bar{\dot{\gamma}}=0.4~$s$^{-1}$ across the flow cell is applied during the whole experiment. Ultrasound is turned on at $t=0$ and switched off at $t=30$~s. Three families of spectra can be distinguished as a function of time before (red), during (blue) and after application of high-power ultrasound (green). The times at which the spectra are shown correspond to the various colored symbols in the inset. The black dash-dotted curve is the form factor reported in blue in Fig.~\ref{fig:saxscb}a. Inset:~Pressure difference $\Delta\Pi$ between the upstream and downstream pressure sensors as a function of time for $P=240$~kPa and $\bar{\dot{\gamma}}=0.4~$s$^{-1}$.
         }
     \label{fig:saxs_flow_hpu}
 \end{figure}

Using the TRUSAXS experiment coupled to high-power ultrasound, we \tg{now} investigate the bulk structure of the gel during flow combined with exposure to ultrasound. \tg{We find that, at least for the relatively small shear rates accessible to the flow cell of our TRUSAXS experiment,} the high-power ultrasound mechanism that leads to the softening of the gel at rest is the same that \tg{facilitates} the flow of the gel: micro-cracks form in the bulk of the gel. 

\begin{figure}
 	\centering
 	\includegraphics[width=1\columnwidth]{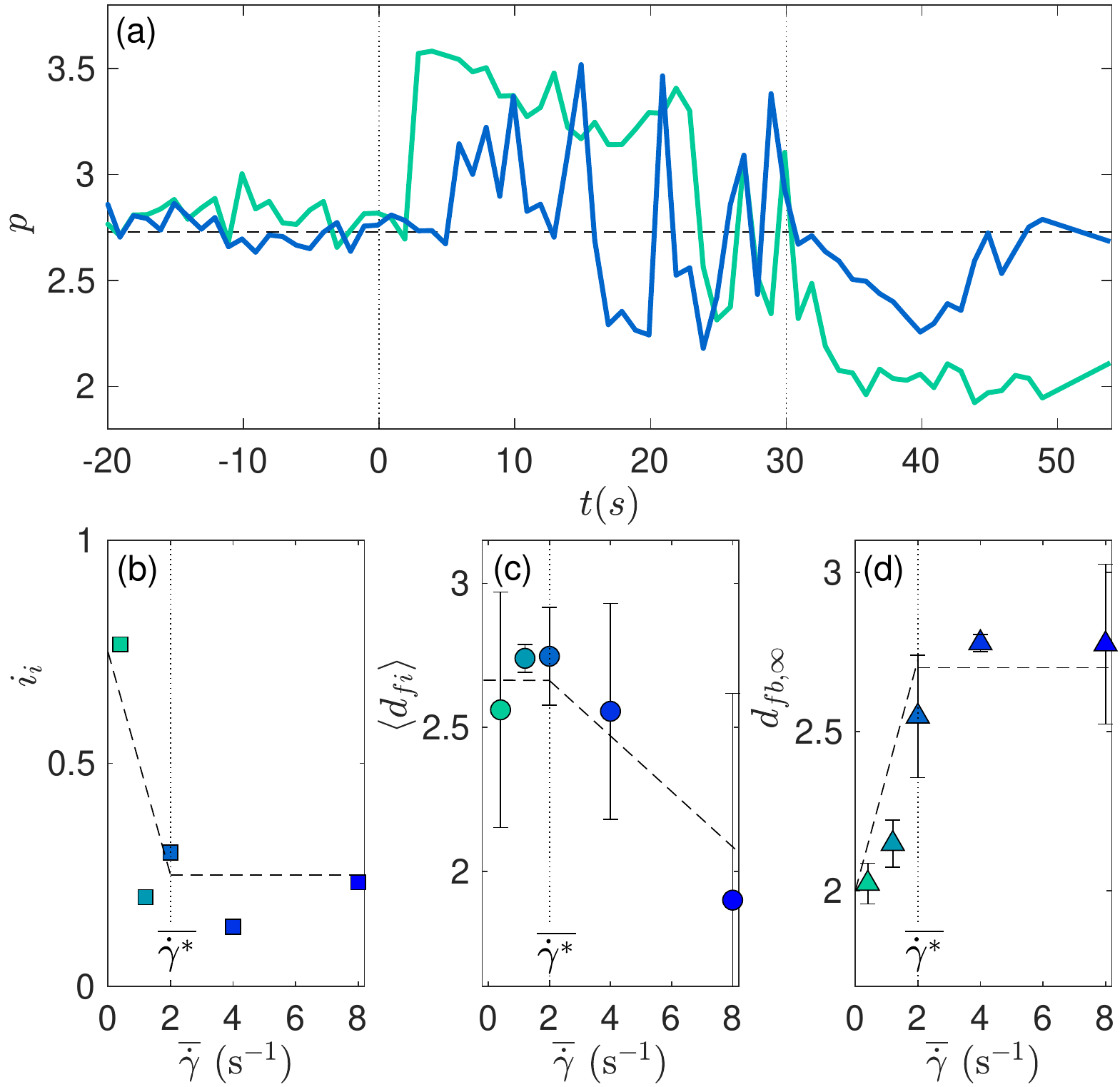}
         \caption{ {\bf Influence of the shear rate on the TRUSAXS spectra of gels exposed to high-power ultrasound.} (a)~Exponent $p$ of the best power-law fit $I(q)\sim q^{-p}$ at low $q$ as a function of time. Ultrasound with frequency $20~$kHz and pressure $P=240~$kPa is turned on at time $t=0$ and switched off at $t=30$~s. A mean shear rate $\bar{\dot{\gamma}}$ across the flow cell is applied during the whole experiment. The green and blue curves correspond to $\bar{\dot{\gamma}}=0.4$ and 2~s$^{-1}$ respectively. (b)~Intermittency index $i_i$ as a function of $\bar{\dot{\gamma}}$. (c)~Micro-cracks average interfacial fractal dimension $\langle d_{fi} \rangle$ as a function of $\bar{\dot{\gamma}}$. (d)~Bulk fractal dimension of the gel $d_{fb,\infty}$ measured 20~s after ultrasound are turned off as a function of $\bar{\dot{\gamma}}$. \tg{For $\dot{\gamma}=4$ and 8~s$^{-1}$, due to the large flow rate imposed by the syringe pump, the gel that passes across the X-ray beam has no longer received any ultrasound at $t\simeq 50$~s, i.e., 20~s after ultrasound are turned off. We therefore measure $d_{fb,\infty}$ on the longest possible time after ultrasound is turned off while the gel has received ultrasound, respectively 4.5~s and 1.5~s for $\dot{\gamma}=4$ and 8~s$^{-1}$.}
          }
     \label{fig:saxs_flow}
 \end{figure}

Thanks to the syringe pump, a constant mean shear rate $\bar{\dot{\gamma}}$ across the flow cell is applied during the whole experiment. Ultrasound with pressure $P$ is turned on at $t=0$~s and switched off at $t=30$~s. TRUSAXS spectra are recorded every second during the entire experiment. As shown in the inset of Fig.~\ref{fig:saxs_flow_hpu} for $P=240$~kPa and $\bar{\dot{\gamma}}=0.4$~s$^{-1}$, ultrasound \tg{facilitates} the flow: we observe a strong drop of the pressure difference $\Delta\Pi$ by about 500~Pa as soon as ultrasound is turned on. For such a low mean shear rate, TRUSAXS spectra are very similar to the ones presented in section \ref{s:structure} where the gel is left at rest. Indeed, the scattering intensity $I(q)$ strongly varies upon application of ultrasound only for $q\lesssim 4\cdot 10^{-3}$~nm$^{-1}$. As shown in Fig.~\ref{fig:saxs_flow}a, the power-law exponent $p$ starts from $p=d_{fb}\simeq 2.75$ \tg{in the presence of flow but} in the absence of ultrasound. \tg{Such a value is similar to the one measured for a gel at rest (see Fig.~\ref{fig:saxs_rest_param_hpu}a)}. \tg{As ultrasound is turned on while the gel flows, $p$} displays values that can be as large as $p=3.5$, which accounts for interfaces of fractal dimension $d_{fi}=6-p\simeq 2.5$. When ultrasound is turned off while the mean shear rate is kept to $\bar{\dot{\gamma}}=0.4$~s$^{-1}$, the intensity spectrum does not fully recover its initial shape and tends to that of a mass fractal with $d_{fb,\infty}=p\simeq \tg{2}$. As explained in section \ref{s:structure}, such a variation of $p(t)$ is compatible with the formation of transient micro-cracks when ultrasound is turned on and with an incomplete structural recovery after ultrasound is turned off. 

Fig.~\ref{fig:saxs_flow}b-d shows the evolution of the various indicators defined in section~\ref{s:structure}, namely $i_i$, $\langle d_{fi} \rangle$ and $d_{fb,\infty}$, as a function of the mean shear rate $\bar{\dot{\gamma}}$ under an acoustic pressure $P=240$~kPa. We may identify two regimes separated by a critical shear rate $\bar{\dot{\gamma}}^*\simeq 2$~s$^{-1}$. With increasing $\bar{\dot{\gamma}}$, $i_i$ decreases and plateaus above $\bar{\dot{\gamma}}^*$: larger shear rates reduce the micro-cracks occurrence formed by high-power ultrasound (Fig.~\ref{fig:saxs_flow}b). In spite of a large experimental uncertainty, $\langle d_{fi} \rangle$ \tg{is consistent with a weak} decreasing trend above $\bar{\dot{\gamma}}^*$, \tg{which may indicate} that micro-cracks interfaces become sharper for larger shear rates. Finally, focusing on the recovery process, a clear increase in the steady-state bulk fractal dimension $d_{fb,\infty}$ is seen from about 2.1 at low shear rates up to a plateau at $d_{fb,\infty}\simeq 2.7$ for $\bar{\dot{\gamma}}>\bar{\dot{\gamma}}^*$. Therefore, highest shear rates make the recovery process complete even after exposure to the highest acoustic pressure achievable with the present TRUSAXS-ultrasound flow cell. To sum up, the scenario depicted at rest remains valid under flow but the effects of high-power ultrasound get smeared out at flow rates larger than $\bar{\dot{\gamma}}^*$.

\subsection{Discussion}

The experiments performed in this section show that high-power ultrasound affects the flow properties of carbon black gels and their microstructure under shear. For shear rates below the characteristic shear rate $\dot\gamma^*\simeq 2.5$~s$^{-1}$ that corresponds to a sharp kink in the flow curve, the effect of ultrasound is much stronger than for $\dot\gamma>\dot\gamma^*$. The most widely accepted interpretation of such a kink in the flow curve is that, upon decreasing the shear rate, $\dot\gamma^*$ corresponds to a transition from a fully fluidized state to a regime dominated by partially arrested flows and slippage at the walls of the shear cell \cite{Meeker:2004b}. The yield stress inferred from a Herschel-Bulkley fit of the low shear branch of the flow curve is generally referred to as a ``pseudo-yield stress'' as it relates to the material properties close to the walls and is dominated by the rheology of the lubricating films at the walls \cite{Seth:2008,Seth:2012}. In the specific case of carbon black gels, it was shown that shear banding and massive wall slip occurs below $\dot\gamma^*$ so that the low shear branch mostly corresponds to a pluglike flow regime while the material is homogeneously sheared along the high shear branch of the flow curve \cite{divoux2013,Radhakrishnan:2017}. Moreover, for such rheopectic materials as carbon black gels, the flow curve is not only highly dependent on the surface properties of the shearing geometry \cite{Grenard:2014} but also on the protocol followed to construct the flow curve \cite{Ovarlez:2013,Helal:2016}. Although complementary local measurements are needed \tg{to investigate the flow field}, it can be hypothesized from the present results that ultrasound promotes wall slip, at least along the low shear branch, and tends to lower both the bulk yield and the ``pseudo-yield stress.''

Furthermore, TRUSAXS experiments reveal that below $\bar{\dot{\gamma}}^*\simeq 2$~s$^{-1}$, the microstructural effects of high-power ultrasound are essentially the same as those observed on the gel at rest. This is consistent with the fact that low shear rates correspond to pluglike flows. Thus, under low shear rates, the gel structure is very similar to the gel structure at rest and micro-cracks occur in the bulk material as well as close to the walls, which is likely to promote wall slip. The fact that $\bar{\dot{\gamma}}^*\simeq \dot\gamma^*$ further strengthens this conclusion, although such a quantitative agreement between the two shear rates might be fortuitous in view of the differences in materials, geometries and protocols. For the largest shear rates, the gel structure is partly rejuvenated by the shear rate and is therefore less sensitive to ultrasound.

\section{Conclusion and outlook}

Thanks to rheological and X-ray scattering measurements performed simultaneously with the application of high-power ultrasound, we have shown that carbon black gels belong to the class of ``rheo-acoustic'' materials, i.e., materials whose mechanical and flow properties can be tuned by ultrasound. More precisely, for acoustic pressures larger than some critical value $P^*$, high-power ultrasound transiently softens and facilitates the flow of carbon black gels through the formation of micro-cracks in the bulk of the gel. \tg{The viscoelastic spectrum at low frequency and the extension of the linear regime are modified even below $P^*$, pointing to ``glass-like'' and more ``fragile'' features.} When shear is applied above a characteristic shear rate $\dot\gamma^*$, the effects of ultrasound on flow properties are mitigated due to shear rejuvenation. Such effects could be further used to fine tune the structure and mechanics of colloidal gels through ultrasonic vibrations in applications where other means of interacting with the microstructure are not possible. Moreover, it is now well established that shear history can be used to tune both linear and nonlinear viscoelastic properties of colloidal gels~\cite{Ovarlez:2013,Koumakis:2015,Helal:2016} and thus to imprint some memory of previous deformations into the material microstructure~\cite{Keim:2013,Mukherji:2019}. Combining these memory effects to the above competition between the effects of shear and of high-power ultrasound will undoubtedly lead to new interesting ways of playing with the microstructure of colloidal gels in search of ``smart'' responses to external stresses.

From a more general point of view, our results still leave some fundamental issues open. A first question concerns the precise microscopic origin of the softening and flow facilitation. We believe that the very large overshoot in the viscous modulus $G''$ upon application of ultrasound and its subsequent relaxation towards a steady-state value about twice larger than the viscous modulus at rest deserve more attention, as it most likely carries key information on the physics underlying ultrasound-induced softening. In particular, strain recovery experiments performed at various stages of the application of high-power ultrasound may help to disentangle the various microscopic contributions to energy dissipation, including viscous drag of the solvent along the gel network, plastic rearrangements and viscoplastic flow of colloidal clusters, and micro-crack formation, in line with the approach proposed recently in Refs.~\cite{Lee:2019,Donley:2020} in the context of yielding under large-amplitude oscillatory shear. 

Furthermore, simultaneous mechanical measurements and microscopic structural characterization under ultrasound are required to provide a full understanding of the effects revealed in the present work. Indeed, at this stage, a joint, quantitative interpretation of both rheological and TRUSAXS experiments is hindered by the differences in setups, protocols and ultrasonic frequencies. Future experiments will include rheo-SAXS measurements under ultrasound to record $I(q)$ spectra simultaneously to rheological data. Moreover, preliminary optical microscopy experiments under ultrasound indicate that micro-cracks may be visualized in carbon black gels, at least once they have grown large enough to lead to detectable contrast variations. Such direct measurements should yield crucial insight into micro-cracks dynamics and into the time scales involved in the transient mechanical response. Finally, coupling small-angle light scattering (SALS) to ultrasound in translucent colloidal gels such as silica gels is also considered in order to test for the generality of the presence of micro-cracks and of their intermittent behaviour in other colloidal gels. Once coupled to both rheometry and ultrasound, time-resolved SALS measurements should provide structural information in quasi-transparent gels on both local and global dynamical processes, similar to TRUSAXS but with much more versatility.

\tg{
\section*{Appendix}
}

\begin{figure}
 	\centering
 	\includegraphics[width=1\columnwidth]{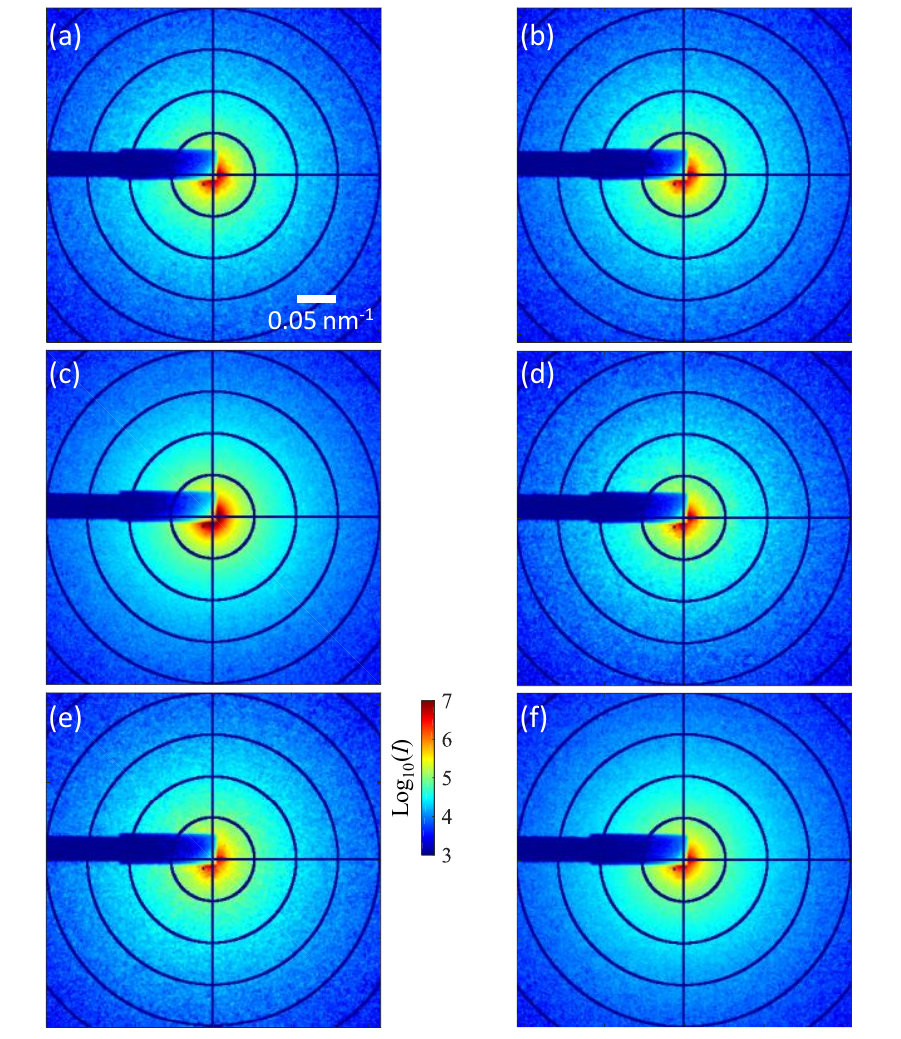}
         \caption{\tg{Two-dimensional TRUSAXS spectra $I(q_x,q_y)$ of the carbon black gel: (a-b)~before, (c-d) during, and (e-f) after application of high-power ultrasound with $P=240$~kPa. Left panels: in the absence of flow $\bar{\dot{\gamma}}=0$~s$^{-1}$. Right panels: under an average shear rate of $\bar{\dot{\gamma}}=2$~s$^{-1}$.
          }
          }
     \label{fig:I2D}
 \end{figure}

\noindent \tg{{\bf Two-dimensional TRUSAXS spectra $I(q_x,q_y)$ --}  Figure~\ref{fig:I2D} shows the evolution of the 2D spectra associated with the experiments presented in Figs.~\ref{fig:saxs_rest_hpu} and \ref{fig:saxs_flow_hpu}. All spectra display radial symmetry, which \seb{indicates} that the carbon black gel keeps an isotropic microstructure during the whole experiments.}

\vspace{.1cm}

\begin{figure}
 	\centering
 	\includegraphics[width=1\columnwidth]{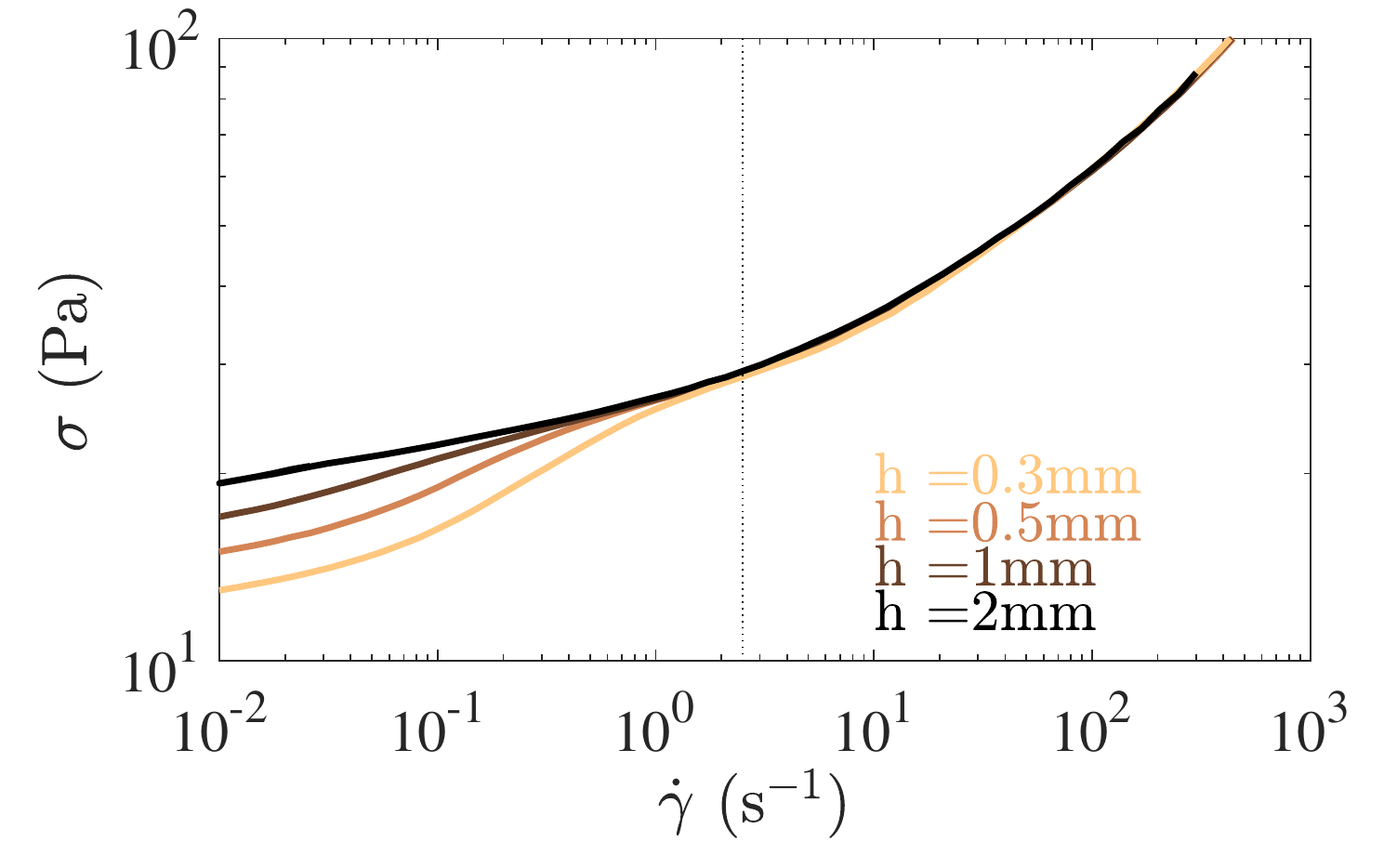}
         \caption{ \tg{Flow curves, shear stress $\sigma$ vs shear rate $\dot\gamma$, of the carbon black gel measured in the absence of ultrasound and in a parallel-plate geometry for different gap widths $h=0.3$, 0.5, 1 and 2~mm from bottom (orange) to top (black). The vertical dashed line shows the characteristic shear rate $\dot\gamma^*\simeq 2.5$~s$^{-1}$ inferred from Fig.~\ref{fig:flowcurve_pierre}.}
          }
     \label{fig:FCgap}
 \end{figure}

\noindent \tg{{\bf Flow curves at different gaps --}  Figure~\ref{fig:FCgap} shows the flow curves measured for four different gaps $h$ in the parallel-plate geometry with a diameter of 35~mm. As in Fig.~\ref{fig:flowcurve_pierre}, the shear rate $\dot\gamma$ is logarithmically swept down from 1000 to 0.01~s$^{-1}$ within $60$~s while the rheometer records the stress response $\sigma$. The flow curves are affected by the gap width below a shear rate $\dot{\gamma}^*$ that strongly depends on $h$, decreasing from about 2 to 0.1~s$^{-1}$ as the gap width is increased. Therefore, instead of observing a plateau indicative of the dynamical yield \seb{stress}, we observe that, for $\dot\gamma<\dot{\gamma}^*$, $\sigma$ decreases all the more with $\dot{\gamma}$ \seb{when} the gap is small. This behavior is typical of a yield stress material that slips at the walls at shear rates below $\dot{\gamma}^*$~\cite{Meeker:2004b,divoux2013,Grenard:2014}. Note that the fact that the flow curves of Fig.~\ref{fig:flowcurve_pierre} lead to a value $\dot{\gamma}^*\simeq 2.5$~s$^{-1}$ that lies in the upper range of the values found here by varying the gap of the parallel-plate geometry can be ascribed to the difference in the surfaces of the shearing tools. While the tool used for Fig.~\ref{fig:flowcurve_pierre} is a smooth cone made out of steel, the upper plate used for the parallel-plate geometry is made out of sandblasted Plexiglas. The former is likely to show stronger wall slip than the latter.}

\section*{Acknowledgements}
The authors thank W.~Chevremont, N.~Hengl, T.~Narayanan, and M.~Sztucki for technical help with the TRUSAXS--ultrasound combined setup at ESRF. Insightful discussions with C.~Barentin, T.~Divoux and A.~Poulesquen on data analysis and interpretation are also acknowledged. This work was funded by the R\'egion Auvergne-Rh\^one-Alpes ``Pack Ambition Recherche'' Programme and by the European Research Council under the European Union’s Seventh Framework Programme (grant agreement No. 258803) and supported by the LABEX iMUST (ANR-10-LABX-0064) of Universit\'e de Lyon, within the program "Investissements d'Avenir" (ANR-11-IDEX-0007) operated by the French National Research Agency (ANR). This work benefited from meetings within the French working group GDR CNRS 2019 “Solliciter LA Mati\`ere Molle” (SLAMM). LRP is part of Labex TEC 21 (Investissements d'Avenir, grant agreement No. ANR-11-LABX-0030), PolyNat Carnot Institute (Investissements d'Avenir, grant agreement No. ANR-11-CARN-030-01) and of IDEX UGA program (ANR-15-IDEX-02).

%

\end{document}